\documentclass[twocolumn,prc,floatfix,preprintnumbers,%
superscriptaddress,nofootinbib]{revtex4}
\usepackage[utf8x]{inputenc}
\usepackage{mathrsfs}
\usepackage{amssymb}
\usepackage{amsmath}
\usepackage{graphicx}

\usepackage{xcolor}
\definecolor{lcolor}{rgb}{0.5,0,0}
\definecolor{citcolor}{rgb}{0,0.3,0.0}

\usepackage[breaklinks,colorlinks,urlcolor=blue,citecolor=citcolor,linkcolor=lcolor]{hyperref}
\usepackage{mciteplus}

\newcommand{\der}{\mathrm{d}}
\newcommand{\rt}{{\mathbf{r}_T}}

\newcommand{\xt}{{\mathbf{x}_T}}
\newcommand{\bt}{{\mathbf{b}_T}}

\newcommand{\yt}{{\mathbf{y}_T}}

\newcommand{\Deltat}{{\boldsymbol{\Delta}_T}}

\newcommand{\tr}{\, \mathrm{Tr} \, }

\newcommand{\nc}{{N_\mathrm{c}}}

\newcommand{\gev}{\ \textrm{GeV}}
\newcommand{\tev}{\ \textrm{TeV}}
\newcommand{\fm}{\ \textrm{fm}}

\newcommand{\as}{\alpha_{\mathrm{s}}}

\def\unitm{\hbox{$1\hskip -1.2pt\vrule depth 0pt height 1.6ex width 0.7pt
	\vrule depth 0pt height 0.3pt width 0.12em$}}

\newcommand{\xpom}{{x_\mathbb{P}}}

\newcommand{\A}{{\mathcal{A}}}

\renewcommand{\Im}{\mathrm{Im}}
\renewcommand{\Re}{\mathrm{Re}}
\begin{document}

\author{Heikki Mäntysaari}
\affiliation{
Department of Physics,  University of Jyväskylä, %
 P.O. Box 35, 40014 University of Jyväskylä, Finland
}
\affiliation{
Helsinki Institute of Physics, P.O. Box 64, 00014 University of Helsinki, Finland
}

\title{
Review of proton and nuclear shape fluctuations at high energy}


\preprint{}

\begin{abstract}
Determining the inner structure of protons and nuclei in terms of their fundamental constituents has been one of the main tasks of high energy nuclear and particle physics experiments. This quest started as a mapping of the (average) parton densities as a function of longitudinal momentum fraction and resolution scale. Recently, the field has progressed to more differential imaging, where one important development is the description of the event-by-event quantum fluctuations in the wave function of the colliding hadron.

In this Review, recent developments on the extraction of proton and nuclear transverse geometry with event-by-event fluctuations from collider experiments at high energy is presented. The importance of this fundamentally interesting physics in other collider experiments like in studies of the properties of the Quark Gluon Plasma is also illustrated.
\end{abstract}

\maketitle

\section{Introduction}

Determining the structure of protons and nuclei in terms of their fundamental constituents and their mutual interactions is one of the long term quests in particle physics. In the last decades, more and more detailed pictures of the (especially) proton structure have been obtained. The most important input for these studies is provided by deep inelastic scattering (DIS) experiments, in which simple pointlike leptons are used to probe the proton substructure. Traditionally, the partonic structure of the proton is presented in terms of the parton distribution functions that describe the quark and gluon densities as a function of longitudinal momentum fraction $x$ of the proton carried by the parton. In particular, thanks to the large amount of precise deep inelastic scattering data from HERA~\cite{Aaron:2009aa,Abramowicz:2015mha}, the quark and gluon distribution functions for the proton are known down to low $x\sim 10^{-5} \dots 10^{-4}$ with good precision.

Beyond the one dimensional distributions (that depend on the longitudinal momentum fraction), it is fundamentally interesting to reveal the geometric structure of proton. In addition to the average size and shape, also the shape fluctuations are basic properties of the proton and should be determined. This knowledge is also crucial for the proper interpretation of many other experimental measurements. The average spatial distribution of partons is encoded in the generalized parton distribution functions~\cite{Mueller:1998fv,Ji:1996nm,Radyushkin:1997ki,Collins:1996fb}.  In this Review, however, a step beyond these distributions is taken and the event-by-event fluctuations of the proton (and nuclear) shape are considered, concentrating on the gluonic structure at small $x$.

There are two different venues that have been recently used to access the geometric shape and shape fluctuations. One can  look at particle correlations in high multiplicity collisions generated  by the (possible) hydrodynamical response to the initial state spatial anisotropies. In heavy ion collisions,  collective phenomena have been used to show the formation fo the Quark Gluon Plasma, and to determine its fundamental properties. In heavy ion collisions the geometry is dominated by the shapes of the colliding nuclei, and the effects sensitive to the nucleon shape (and shape fluctuations) are less important. In proton-lead and proton-proton collisions the initial state geometry is controlled by the single proton, and the shape fluctuations can potentially be observed more clearly assuming that a hydrodynamically evolving fluid is produced. However, the applicability of hydrodynamics when describing such a small systems produced in proton-proton or proton-lead collisions is not completely established and is currently under an intensive research.

An another approach is to look at observables that  directly probe the event-by-event fluctuations in the target wave function. As  will be discussed in detail in this Review, one of these processes is incoherent diffraction in deep inelastic scattering. In these events, the target proton or nucleus dissociates and a rapidity gap exists between the target remnants and the diffractively produced system (e.g. a vector meson).  In addition, one can consider fully elastic scattering processes to obtain more indirect evidence of the proton shape fluctuations.

In this review the focus in on the proton and nuclear shape fluctuations at high energy (small momentum fraction $x$).
In Sec.~\ref{sec:diffraction} it is shown how diffractive processes can be used to constrain the average shape of the target, as well as probe the event-by-event fluctuations at different distance scales. A  theoretical framework used to describe these high energy processes based on the Color Glass Condensate effective theory of QCD is shortly introduced in Sec.~\ref{sec:dipole}. Based on this discussion, in Sec.~\ref{sec:diffraction_analysis} we review the recent progress in using the diffractive vector meson production to constrain the proton shape fluctuations. Elastic proton-proton scattering, and particle correlations in high-multiplicity proton-proton collisions, are discussed in the context of fluctuating proton structure in Sec.~\ref{sec:pp}. Fluctuating nucleons in nuclei in the context of diffractive vector meson production and hydrodynamical simulations of the Quark Gluon Plasma production in proton-lead collisions are reviewed in Sec.~\ref{sec:nuclei}.

\section{Exclusive scattering as a probe of transverse geometry}
\label{sec:diffraction}

As discussed above, the total cross section in DIS can be studied to measure the total partonic densities in protons and nuclei. To obtain a more detailed picture of the target, more differential observables are needed. One especially powerful process is exclusive production of a vector meson with the same quantum numbers $J^{PC}=1^{--}$ as the photon in deep inelastic scattering:
\begin{equation}
\gamma^* + p/A \to V + p/A.
\end{equation}
Here $V$ refers to a vector meson ($V=\rho,\phi,J/\psi, \Upsilon,\dots$), and the target can be a proton $p$ or a nucleus $A$. An experimental signature is a rapidity gap (empty detector) between the produced vector meson and the target. The target can either remain in the same quantum state or dissociate, these processes will be later considered separately. In general, these processes are called diffractive.

There are two main advantages that make vector meson production powerful. First, in order to produce the vector meson and nothing else, there can not be a net color charge transfer to the target. This requires at least two gluons to be exchanged with the target, and the cross section consequently scales approximatively as the gluon density squared~\cite{Ryskin:1992ui}.
The other advantage is that only in the exclusive scattering process it is possible to measure the total momentum transfer $\Deltat$, and interpret that as the Fourier conjugate  to the impact parameter. Consequently, these processes probe not only the density of partons, but also their spatial distribution.

Let us first consider coherent diffraction, in which the target remains in the same quantum state after the interaction. In this case, the cross section is calculated by applying the Good-Walker formalism~\cite{Good:1960ba}, which states that the cross section is determined by the average interaction of the states that diagonalize the scattering matrix with the target. At high energies, these states are the Fock states of the incoming virtual photon with a fixed number of partons (at leading order a quark-antiquark pair) at fixed transverse coordinates, probing a fixed configuration of the target. This formalism can then be generalized to also include fluctuating target configuration. The cross section then reads~\cite{Miettinen:1978jb} (following the notation of Ref.~\cite{Kowalski:2006hc})
\begin{equation}
\label{eq:coherent}
\frac{\der \sigma^{\gamma^*A \to VA}}{\der t} = \frac{1}{16\pi} \left| \left \langle \mathcal{A}^{\gamma^* A \to VA} \right \rangle \right|^2,
\end{equation}
where $A$ refers to the target proton or nucleus,  $t$ is the squared four momentum transfer (at high energies $t \approx -\Deltat^2$) and $\mathcal{A}$ is the  amplitude for the diffractive scattering (the same amplitude also enters in the calculation where the target dissociates, as demonstrated shortly).  
As the average is taken over all possible configurations of the incoming proton or nucleus, the scattering amplitude is sensitive to the average interaction of the quark dipole with the target. This, in turn, is controlled by the average distribution of small-$x$ gluons in the target wave function. Thus, the coherent cross section generally probes the average structure of the target. 
We will return to the calculation of $\mathcal{A}$ later in Sec.~\ref{sec:dipole}.

Let us then consider the class of events in which the outgoing proton or nucleus is not in the same quantum state as the incoming particle following Ref.~\cite{Caldwell:2009ke}. These events are called incoherent, and the experimental manifestation is that the target dissociates into a system of particles, maintaining the rapidity gap to the produced vector meson. To describe these events, the scattering amplitude in the case where initial state is $| i\rangle$ and the final state is $|f\rangle$ is expressed as
$\left \langle f \left|   \mathcal{A}   \right| i \right \rangle$ (we use a shorthand notation $\mathcal{A}$ to refer to the diffractive scattering amplitude $\mathcal{A}^{\gamma^* A \to VA}$). The initial and final states are required to be different. The squared transition amplitude, which enters in the cross section, reads
\begin{multline}
\label{eq:incoh_states}
\sum_{f\neq i} \left| \left \langle f \left|   \A   \right| i \right \rangle \right|^2 =  \sum_f  \langle i | \A^* | f \rangle \langle f | \A | i \rangle   - \left\langle i \left | \A \right | i \right \rangle \left\langle i \left | \A^* \right | i \right \rangle \\
= \langle i | \A^*\A | i \rangle - |\langle i | \A | i \rangle|^2.
\end{multline}
To get the last line the completeness of states, $\sum_f |f\rangle \langle f | = \unitm$, is used. Note that the sum over final states includes all possible states for the final state proton, and not only the different spatial configurations of its ground state. To get the cross section, the squared amplitude must be averaged over the possible initial states $|i\rangle$. The first term in Eq.~\eqref{eq:incoh_states} then becomes the squared diffractive scattering amplitude  $\mathcal{A}^{\gamma^* A \to VA}$ averaged over the initial configurations, and the second term gives the average of the diffractive scattering amplitude squared. The resulting incoherent cross section then reads
\begin{multline}
\label{eq:incoh_final}
\frac{\der \sigma^{\gamma^*A \to VA^*}}{\der t}  =\frac{1}{16\pi} \left( \left \langle \left| \A^{\gamma^*A \to VA} \right|^2 \right\rangle \right. \\
\left. - \left| \left \langle \A^{\gamma^*A \to VA} \right\rangle \right|^2 \right).
\end{multline}
Here $A^*$ denotes the final state proton or nucleus in a state which differs from the initial state. 
Due to the different ordering of the squaring and averaging of the amplitude $\A^{\gamma^*V \to VA}$, the incoherent cross section measures how much the scattering amplitude fluctuates between the different possible initial state configurations. Because the total momentum transfer $\Deltat \approx \sqrt{-t}$ is measurable and the Fourier conjugate to the impact parameter, this property can be used to constrain the geometry fluctuations in the target at different length scales as demonstrated shortly. 
When applying Eq.~\eqref{eq:incoh_final} away from $|t|\ll 1/R_A^2$ where it is strictly applicable~\cite{Frankfurt:1994hf} ($R_A$ being the target size), a specific model for the fluctuating structure in the transverse plane is needed, as originally presented in Ref.~\cite{Miettinen:1978jb}. Additionally, one has to assume that the non-zero momentum transfer does not mix the states with different transverse structure that are used to diagonalize the scattering matrix.
 Up to how large $|t|$ this extension is valid is not yet completely determined.

An analogous process to the exclusive vector meson production is deeply virtual compton scattering (DVCS), in which a real photon is produced. At high energies where the target is dominated by gluons, the DVCS process is also sensitive to the small-$x$ gluons and their spatial distribution in the target. The  difference to the exclusive vector meson production is that one does not need to model the non-perturbative vector meson formation process, and the mass of the meson does not provide a hard perturbative scale. On the other hand, one can also work in terms of the generalized parton distribution functions or GPDs (see Ref.~\cite{Diehl:2003ny} for a review), that depend both on the longitudinal momentum fraction and transverse coordinate. For the possibilities to use DVCS to constrain the quark and gluon GPDs in the future Electron-Ion Collider,  see e.g. Ref.~\cite{Accardi:2012qut}.

To summarize, the diffractive processes  provide  unique constraints on the target structure. First, the coherent cross section measures  the average parton distribution. In exclusive scattering where the impact parameter dependence is probed, coherent vector meson production gives access to the spatial distribution of small-$x$ partons (dominantly gluons).
Simultaneously, incoherent vector meson production  where the target dissociates measures the degree of fluctuations in the scattering amplitude (at the length scale controlled by the inverse momentum transfer) is measured.  In Sec.~\ref{sec:diffraction_analysis} it is show how the proton geometry fluctuations can be obtained by applying these two constrains simultaneously\footnote{In these analyses no constraints on the impact parameter are imposed. However, as proposed in Ref.~\cite{Lappi:2014foa}, it may also be possible to analyze incoherent diffraction at different centrality classes.}.

\section{Dipoles probing protons and nuclei at high energy}
\label{sec:dipole}
\begin{figure}[tb]
		\includegraphics[width=\columnwidth]{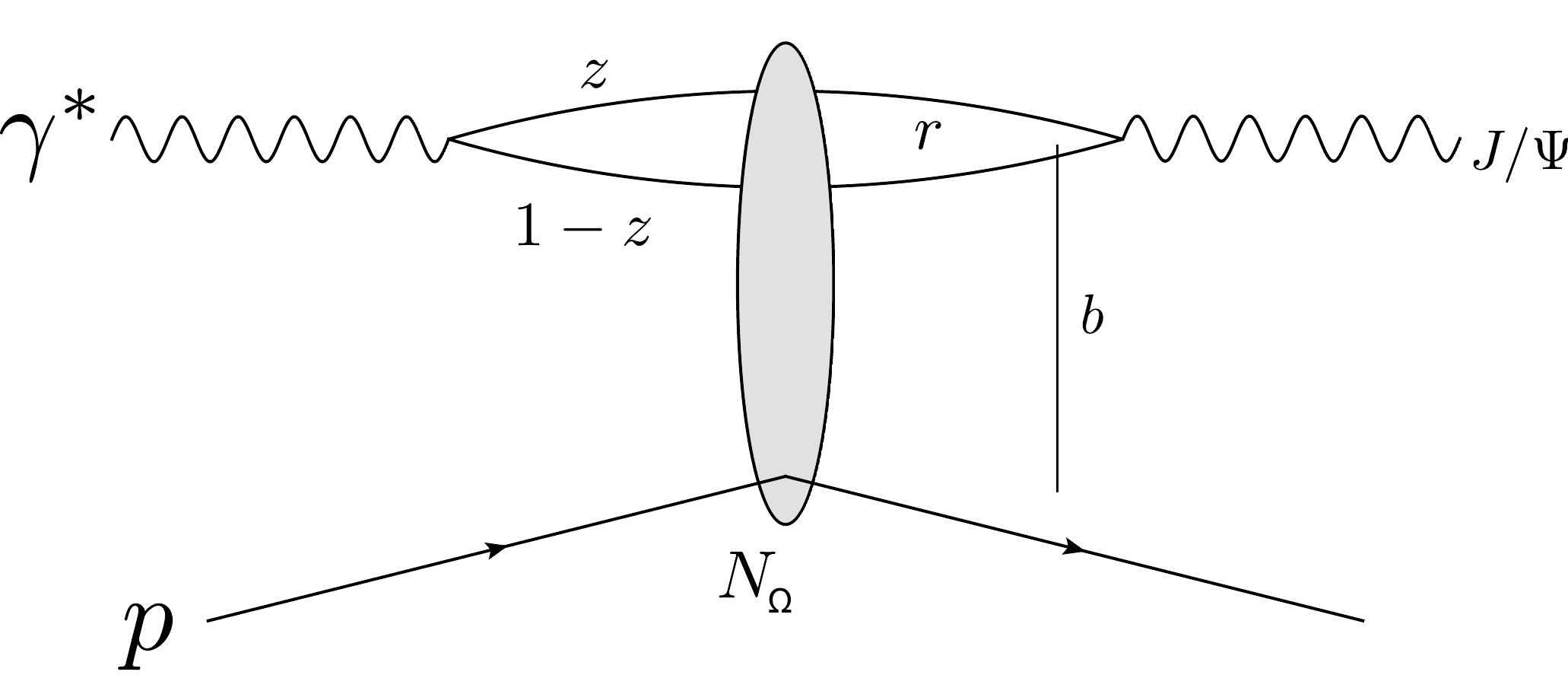} 
				\caption{Exclusive J/$\psi$ production in virtual photon-proton scattering. The grey blob refers to the multiple scattering in the target color field, described by the dipole scattering amplitude $N_\Omega$. }  
			\label{fig:jpsi_diag}
\end{figure}

At high energies, the lifetime of the quark-antiquark fluctuation is much longer than the timescale of the dipole-target interaction. Consequently, the production cross section factorizes, and it becomes natural to describe the scattering processes in the dipole picture~\cite{Mueller:1993rr,Nikolaev:1991et}. 
 First, the incoming photon splits to a quark-antiquark dipole, which is a simple QED process~\cite{Kovchegov:2012mbw}. Subsequently, the dipole scatters elastically off the target. Long after the interaction, non-perturbative vector meson formation from the quark dipole takes place. This factorization makes it possible to write the diffractive scattering amplitude in the Color Glass Condensate framework (for a review, see e.g. Refs.~\cite{Iancu:2003xm,Gelis:2010nm,Albacete:2014fwa,Blaizot:2016qgz}) as~\cite{Kowalski:2006hc}
\begin{multline}
\label{eq:diffamp}
\A^{\gamma^*A \to VA} = 2 i \int \der^2 \rt  \frac{\der z}{4\pi} \der^2 \bt e^{-i\bt \cdot \Deltat - (1-z)\rt \cdot \Deltat} \\
   \times  \Psi^*(\rt,z,Q^2) \Psi_V(\rt,z,Q^2) N_\Omega(\rt,\bt,\xpom)
\end{multline}
illustrated in Fig.~\ref{fig:jpsi_diag}.
Here, the virtual photon wave function $\Psi$ describes the photon to dipole splitting, and its overlap with the vector meson wave function $\Psi_V$ controls the contribution of different dipole sizes $\rt$ and longitudinal momentum fractions $z$ at fixed virtuality $Q^2$. The dipole-target scattering is described by the dipole amplitude $N$, which depends on the particular configuration of the target $\Omega$. The transverse momentum transfer $\Deltat$ is the Fourier conjugate to the impact parameter (center of the dipole) $\bt$ as discussed earlier, with the correction $(1-z)\rt \cdot \Deltat$ taking into account that the vector meson wave function is not taken in the forward kinematics~\cite{Bartels:2003yj}\footnote{It has recently been argued in Ref.~\cite{Hatta:2017cte} that the conjugate to the momentum transfer should be $\bt + \left(\frac{1}{2}-z\right)\rt$, instead of $\bt + (1-z)\rt$. However, numerical results are almost identical in both cases.}. The target structure is probed at the longitudinal momentum fraction
\begin{equation}
\xpom = \frac{M_V^2 + Q^2-t}{W^2+Q^2-m_N^2},
\end{equation}
which is analogous to the Bjorken-$x$ of DIS. Here $W$ is the center-of-mass energy for the photon-nucleon scattering, $M_V$ the vector meson mass, $Q^2$ the photon virtuality and $m_N$ the  nucleon mass.

The connection to the transverse geometry is apparent in Eq.~\eqref{eq:diffamp}, as the dipole-target scattering amplitude $N_\Omega$ is Fourier transformed from the coordinate space to the momentum space. When the coherent cross section is calculated, one first takes the average of $\A^{\gamma^*A \to VA}$ over different target configurations. This averaging results in average dipole-target scattering amplitude $N(\rt,\bt,\xpom) = \langle N_\Omega (\rt,\bt,\xpom)\rangle$. As the dipole amplitude is approximatively proportonal to the gluon density at the given impact parameter $\bt$, 
the $t$ distribution of the  coherent cross section is approximatively the 2-dimensional Fourier transform of the spatial gluon distribution. The extraction of the impact parameter dependence of the dipole amplitude was discussed for the first time in Ref.~\cite{Munier:2001nr} (see also Ref.~\cite{Ralston:2001xs}), inspired by the determination of the proton density profile based on elastic proton-proton collision measurements presented in Ref.~\cite{Amaldi:1979kd}.

It is customary to include two corrections in phenomenological studies of exclusive vector meson production following Ref.~\cite{Kowalski:2006hc}. First, one usually assumes that the dipole  amplitude is purely real and consequently the vector meson production amplitude \eqref{eq:diffamp} imaginary. By requiring that the scattering amplitude is an analytical complex function, the real part can be recovered and included as a numerically small (generally $\sim 10\%$ at the cross section level) correction. Additionally, the so called skewedness correction which takes into account the fact that in the two-gluon exchange limit the two gluons carry a different fraction of the target longitudinal momentum~\cite{Shuvaev:1999ce}. This correction is numerically large (generically up to $\sim 40\%$, see e.g. Ref.~\cite{Mantysaari:2017dwh}), and its applicability in the dipole picture discussed here where the two quarks undergo multiple scattering in the target color field is not completely established.

A major advantage of the CGC approach is that both inclusive and diffractive observables can be described in the same framework in terms of same degrees of freedom. This is not the case in collinear factorization approach, as the parton distribution functions are by definition inclusive, and relating them to exclusive processes is not straightforward (see however~\cite{Guzey:2013qza}). In CGC, the QCD dynamics is described in terms of the dipole-target scattering amplitude $N$ which includes all information about the target structure. As an illustration, let us consider Eq.~\eqref{eq:diffamp} in the case where the final state is a virtual photon, and write $\Psi_V = \Psi$ where $\Psi$ is the photon wave function. Now, the optical theorem states that the total photon-proton cross section is related to the imaginary part of the forward scattering amplitude ($\Deltat=0$). As the dipole amplitude $N$ is generally assumed to be real and the cross section is obtained by averaging over all possible target configurations, the total photon-proton (or photon-nucleus) cross section becomes
\begin{multline}
\sigma^{\gamma^*p } = \Im \A^{\gamma^*A \to \gamma^*A}(\Deltat=0)  \\=  2 \int \der^2 \rt  \frac{\der z}{4\pi} \der^2 \bt |\Psi(\rt,z,Q^2)|^2 \langle N(\rt,\bt,\xpom) \rangle.
\end{multline}
We immediately note that unlike diffractive cross sections, inclusive cross section entering e.g. in the structure functions is only sensitive to the gluon density integrated over the whole target. As such, it can not distinguish between different impact parameter distributions of the same overall gluon density. In general, determining the collision centrality in deep inelastic scattering is challenging, see however Ref.~\cite{Zheng:2014cha} where the forward neutron multiplicity is proposed to act as a proxy for the centrality in the future Electron Ion Collider. 

In the Color Glass Condensate framework, the dipole-target scattering amplitude satisfies perturbative evolution equations, such as the JIMWLK~\cite{JalilianMarian:1996xn,JalilianMarian:1997jx,JalilianMarian:1997gr,Iancu:2001md} equation (named aver Jalilian-Marian, Iancu, McLerran, Weigert, Leonidov and Kovner). This equation, or its NLO version~\cite{Balitsky:2013fea,Kovner:2013ona}, is equivalent to the infinite hierarchy or coupled differential equations known as the Balitsky hierarchy, which  reduces to a closed equation in the  mean field and large-$\nc$ limit, known as the Balitsky-Kovchegov (BK) equation~\cite{Kovchegov:1999yj,Balitsky:1995ub} also available at the NLO accuracy~\cite{Balitsky:2008zza,Balitsky:2013fea}. The initial condition, describing e.g. the dipole amplitude $N$ at the initial Bjorken-$x$, is a non-perturbative input to the evolution and is usually obtained by fitting the HERA deep inelastic scattering data as in Refs.~\cite{Albacete:2010sy,Lappi:2013zma,Ducloue:2019jmy}. A challenge in this approach is to describe the evolution of the target geometry, as the perturbative evolution equations produce long-distance Coulomb tails that result in cross sections growing much faster than seen in the measurements. This growth should be tamed by confinement scale effects, whose implementation is challenging, see e.g.~\cite{Berger:2011ew,Mantysaari:2018zdd}. Consequently, in phenomenological applications one often applies parametrizations for the dipole amplitude that include dependence on the proton geometry, and are  matched to perturbative QCD in the dilute region. A common choice is the IPsat parametrization~\cite{Kowalski:2003hm} with free parameters fitted to HERA data in~\cite{Rezaeian:2012ji,Mantysaari:2018nng}. In this parametrization the dipole amplitude is written as
\begin{equation}
\label{eq:ipsat}
N(\rt,\bt,x) = 1 -\exp \left( -\rt^2 F(\rt,x) T_p(\bt) \right) 
\end{equation}
with
\begin{equation}
F(\rt,x)=\frac{\pi^2}{2 \nc} \as(\mu^2) xg(x, \mu^2).
\end{equation}
Here $T_p$ is the proton density profile with possible event-by-event fluctuations (normalized such that $\int \der^2 \bt T_p(\bt)=1$), and $xg$ the gluon distribution which satisfies the DGLAP equation~\cite{Gribov:1972ri,Gribov:1972rt,Altarelli:1977zs,Dokshitzer:1977sg} to guarantee a smooth matching to perturbative QCD at small dipoles $\rt$ corresponding to the dilute region.

Finally, to conclude this Section let us briefly mention the connection to the collinear factorization.  In the non-relativistic limit, where the quark and antiquark carry the same longitudinal momentum ($z=0.5$) and when the multiple interactions with the target are neglected, one recovers the so called \emph{Ryskin result}~\cite{Ryskin:1992ui} for the coherent cross section as shown explicitly in Ref.~\cite{Anand:2018zle}:
\begin{equation}
\frac{\der \sigma_{\rm coh}}{\der t} = \frac{\pi^3 \alpha_s^2 M_V^3 \Gamma_{V \to e^+e^-}}{3 \alpha_\text{em}} \left[ \frac{1}{(2\bar q^2)^2}   xg( x, \bar q^2)\right]^2 F_N(t)^2.
\end{equation}
In this form the sensitivity on the small-$x$ gluons is apparent, as the cross section is proportional to the squared gluon density $xg$ (although the skewedness correction discussed above could be applied), and the $t$ dependence is parametrized in terms of the target form factor $F_N$. However, this simple treatment neglects the fact that the inclusive structure functions can not be directly used when describing exclusive processes. Here $\Gamma_{V\to e^+e^-}$ is the leptonic decay width of the vector meson $V$ having mass $M_V$, and $\bar q = M_V^2/4$ the momentum scale of the process.

\section{Proton structure fluctuations from diffraction}
\label{sec:diffraction_analysis}
\subsection{Constraining the fluctuations at small $x$}
\label{sec:fluct_fixed_x}
In Sec.~\ref{sec:diffraction} it was  discussed how the coherent cross section is sensitive to the average structure of the target, and how the incoherent cross section measures the amount of event-by-event fluctuations. In this section, the recent progress in constraining the fluctuating transverse geometry of the proton based on HERA vector meson production data is reviewed.

Coherent and incoherent vector meson production has been extensively measured at HERA (where the incoherent process is usually called proton dissociation), where electrons and positrons were collided with protons at center-of-mass energy $\sqrt{s}=318\gev$. This translates into photon-proton center-of-mass energies $W$ in the range $W=30 \dots 300 \gev$. Both H1 and ZEUS measured diffractive $J/\psi$ production in the photoproduction ($Q^2\lesssim 1\gev^2$) and electroproduction ($Q^2\gtrsim 1 \gev^2$) regions~\cite{Breitweg:1999jy,Chekanov:2002xi,Aktas:2003zi,Chekanov:2004mw,Aktas:2005xu,Chekanov:2009ab,Alexa:2013xxa}. In addition to $J/\psi$, exclusive production of lighter $\rho$~\cite{Adloff:1999kg,Aid:1996ee,Chekanov:2007zr,Aaron:2009xp} and $\phi$~\cite{Chekanov:2005cqa,Aaron:2009xp} mesons as well as heavier upsilons~\cite{Breitweg:1998ki,Chekanov:2009zz,Adloff:2000vm} has also been measured.

The possibility to constrain the event-by-event fluctuations in the proton was pioneered in Ref.~\cite{Mantysaari:2016ykx} based on ideas discussed in Refs.~\cite{Miettinen:1978jb,Caldwell:2009ke}, first by using the IPsat parametrization, Eq.~\eqref{eq:ipsat}, to describe the dipole-proton interaction. In this case, if there are no fluctuations in the proton density profile $T_p(\bt)$,  the incoherent cross section \eqref{eq:incoh_final} vanishes. On the other hand, the coherent spectrum is sensitive to the proton profile. As the  mass of the vector meson suppresses contributions from large dipoles $|\rt| \gtrsim 1/M_V$, one can approximatively linearize the IPsat dipole \eqref{eq:ipsat}, substitute that  to the diffractive scattering amplitude \eqref{eq:diffamp} and obtain an estimate for the $t$ dependence of the coherent cross section:
\begin{equation}
\label{eq:coh_tslope}
\frac{\der \sigma^{\gamma^* p \to V p}}{\der t} \sim e^{-B_p |t|},
\end{equation}
when the proton density profile is assumed to be Gaussian:
\begin{equation}
T_p(\bt) = \frac{1}{2\pi B_p} e^{-\bt^2/(2B_p)}.
\end{equation}
Thus, the $t$ slope directly measures the size of the proton: the transverse proton radius is $\sqrt{2B_p}$.
In practice, as the impact parameter profile is more complicated in the IPsat parametrization, the Fourier transform results in diffractive minima at large $|t|$ (outside  the range where H1 and ZEUS were able to measure the spectra), see discussion in Ref.~\cite{Armesto:2014sma}. This is a general feature in diffractive scattering, and the location of the first minimum is controlled by the size of the target $R$ as  $t_\text{first dip} \sim 1/R^2$.

To obtain also a  non-zero incoherent cross section, the proton density profile $T_p$ was taken to have event-by-event fluctuations in Ref.~\cite{Mantysaari:2016ykx}. The simplest ansatz, motivated by the picture where the three constituent quarks emit small-$x$ gluons around them, is to assume that the proton consists of $N_q$ hot spots at random locations $\bt_i$ (see also Ref.~\cite{Mitchell:2016jio} for a more detailed discussion of possible ways to sample constituent quark structure). In this case, the density profile is written as
\begin{equation}
\label{eq:hotpsottp}
T_p(\bt) \to \frac{1}{N_q} \sum_{i=1}^{N_q} \frac{1}{2 \pi B_q} e^{-(\bt-\bt_i)^2/(2B_q)}.
\end{equation}
In a simple case, inspired by the constituent quark model, one can take $N_q=3$, but  other choices are also possible~\cite{Mantysaari:2016jaz,Cepila:2016uku}. The distances for the hot spots from the center of the proton are sampled from a Gaussian distribution whose width $B_{qc}$ is a free parameter, as is the size of the hot spot  $B_{q}$. Requiring a simultaneous description of both coherent and incoherent $J/\psi$ production data, optimal values for these parameters can be determined. 

\begin{figure}[tb]
		\includegraphics[width=\columnwidth]{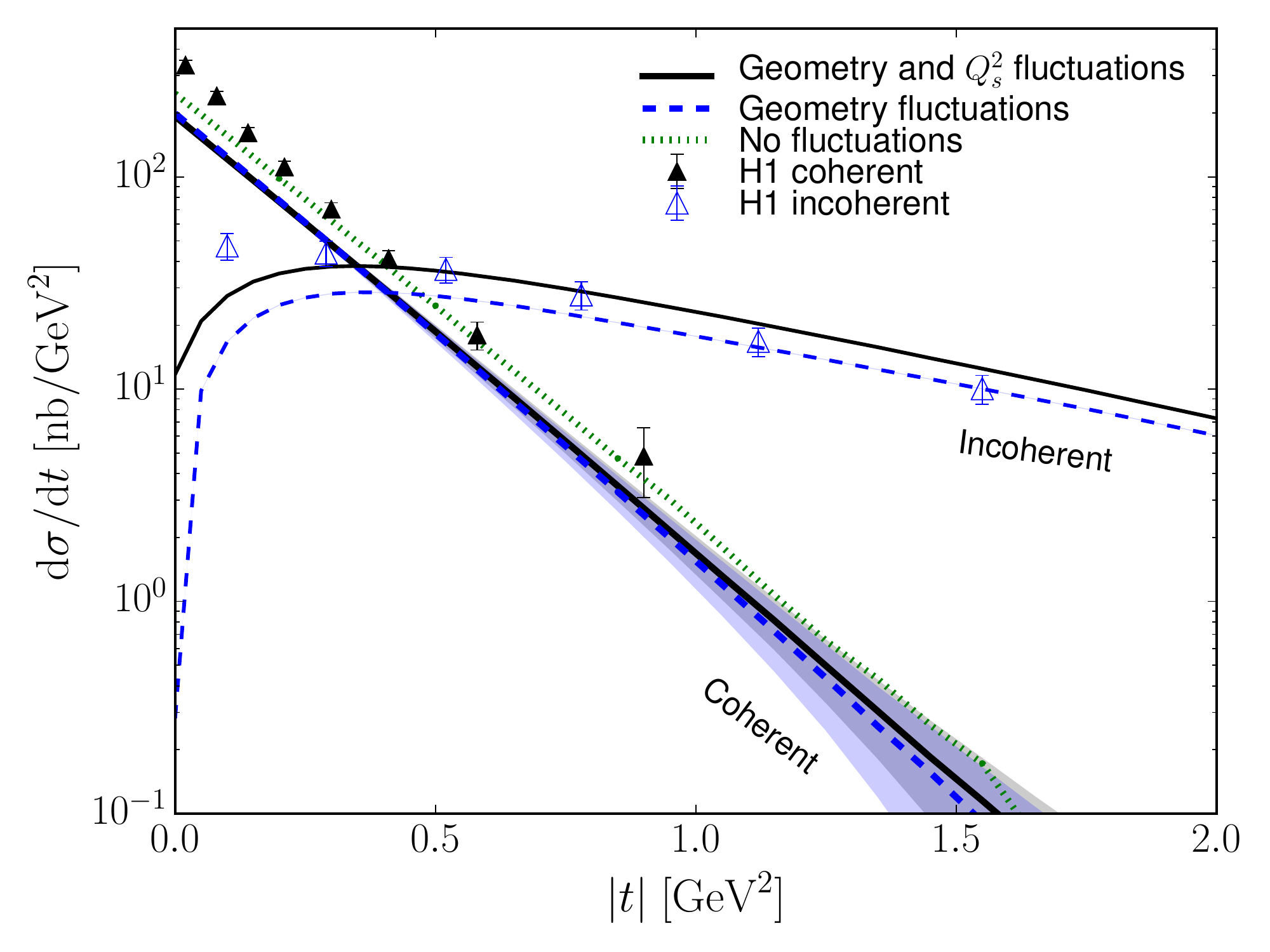} 
				\caption{$J/\psi$ photoproduction cross sections at $W=75\gev$ as a function of squared momentum transfer as measured by H1~\cite{Aktas:2005xu}, compared to calculations using the IPsat parametrization for the dipole-target scattering. Geometric shape fluctuations and overall normalization fluctuations ($Q_s^2$ fluctuations) are needed to describe the data. Figure based on Ref.~\cite{Mantysaari:2016jaz}. }  
			\label{fig:ipsat_fluct_nofluct}
\end{figure}

Coherent and incoherent $J/\psi$ photoproduction ($Q^2=0$)  spectra computed using a spherical proton (in which case the incoherent cross section vanishes) and a fluctuating proton with the optimal parameters $B_q=0.7\gev^{-2}$ and $B_{qc}=3.3\gev^{-2}$ from Ref.~\cite{Mantysaari:2016jaz} are shown in Fig.~\ref{fig:ipsat_fluct_nofluct}. Note that the Bjorken-$x$ probed here is approximatively $\xpom \approx 10^{-3}$.  As the characteristic hot spot separation $B_{qc}$ turns out to be much larger than the hot spot size $B_q$, large spatial geometry fluctuations are needed to obtain a good fit to the HERA data. 
Although the structure fluctuations mostly affect the incoherent cross section (which would otherwise vanish in this calculation), the coherent cross section is also somewhat modified by the substructure fluctuations. As the coherent cross section is only sensitive to the average structure of the target, or to the average dipole-target scattering amplitude, it would also be possible in principle to construct a parametrization for the fluctuating proton structure which leaves the average dipole-target interaction intact. Note that Eq.~\eqref{eq:hotpsottp} modifies the average proton shape, due to the non-linear dependence on the density function $T_p$  in the dipole amplitude~\eqref{eq:ipsat}. Consequently, the different results for the coherent cross section should not be taken to quantify the effect of proton shape fluctuations on coherent J/$\psi$ production.
 

At large $|t|$ the slope of the incoherent spectrum is controlled by the size of the smallest object that fluctuates as shown explicitly in Ref.~\cite{Lappi:2010dd}. At small $|t|$, on the other hand, one is sensitive to the fluctuations at long distance scale $\sim 1/\sqrt{|t|}$. With only geometry fluctuations included in the model, the small-$|t|$ part of the incoherent spectra is underestimated. To improve the description of the HERA data, in Ref.~\cite{Mantysaari:2016ykx} additional density fluctuations were implemented following Ref.~\cite{McLerran:2015qxa}, by allowing the density of each of the hot spots to fluctuate independently. As can be seen in Fig.~\ref{fig:ipsat_fluct_nofluct}, when the density (labeled as $Q_s^2$) fluctuations are included on top of geometry fluctuations, an improved description of the HERA data is obtained at all $|t|$.

It is crucial to note that the hot spot picture applied above is only one possible fluctuating density profile for the proton compatible with the HERA vector meson production data. The incoherent cross section measures how large fluctuations there are in the diffractive scattering amplitude, and the distance scale probed is set by $\sim 1/|t|$. There are multiple different fluctuating geometries that result in the same variance. This was explicitly demonstrated in Ref.~\cite{Mantysaari:2016jaz} where instead of a hot spot structure, the proton density profile was constructed based on a lattice QCD inspired picture where the constituent quarks are connected via color flux tubes that merge at a specific point inside the proton~\cite{Bissey:2006bz}. Fixing again the free parameters, the tube width and the width of the distribution from which the constituent quark positions are sampled, it is possible to obtain an equally good description of the coherent and incoherent $J/\psi$ production data. This means that the average proton shape and the amount of fluctuations are approximatively the same as in the hot spot parametrization.

In addition to geometry and density fluctuations, one expects to  have individual color charge fluctuations in the proton. To estimate their contribution to the incoherent cross section, one needs to go beyond the IPsat parametrization. One possible approach is to apply the McLerran-Venugopalan model in the CGC framework~\cite{McLerran:1993ni}. Here, the fluctuating color charges are included by assuming that the color charges are local random Gaussian variables, with the width of the distribution proportional to the local density, characterized by the saturation scale $Q_s^2$. The color charge density $\rho^a$ (where $a$ is the color index) in the MV model is thus sampled from a distribution
\begin{equation}
\langle \rho^a(x^-,\xt) \rho^b(y^-,\yt) \rangle =  \delta^{ab} \delta^{(2)}(\xt-\yt) \delta(x^- - y^-) g^2 \mu^2
\end{equation}
with the expectation value zero. The color charge density $\mu^2$ is taken to be directly proportional to the local  saturation scale $Q_s^2(\bt)$ which can be extracted from the IPsat parametrization. The geometry and density fluctuations can thus be included by introducing them in the IPsat parametrization as discussed above. 

In practice, in this approach one samples the color charges for each event, and then solves the classical Yang-Mills equations to obtain the Wilson lines $V(\xt)$ at each point on the  transverse plane. The Wilson lines describe the color rotation of a high energy quark when it propagates through the target at fixed transverse coordinate $\xt$, and the dipole-target amplitude is obtained as a trace of  two Wilson lines: 
\begin{equation}
N_\Omega (\rt,\bt) = 1- \frac{1}{\nc} \tr V^\dagger(\bt-\rt/2) V(\bt + \rt/2).
\end{equation}

\begin{figure}[tb]
		\includegraphics[width=\columnwidth]{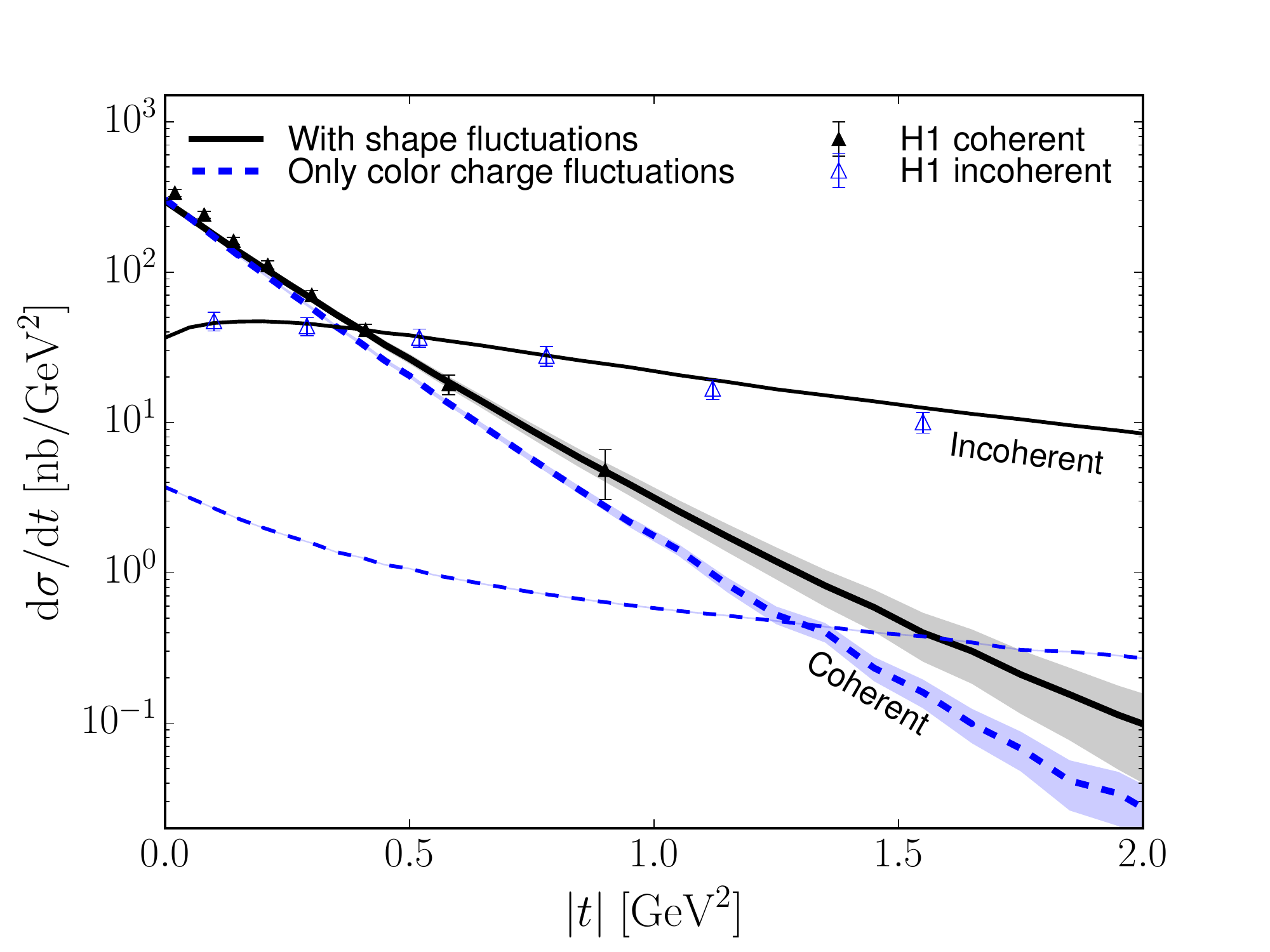} 
				\caption{Coherent and incoherent $J/\psi$ photoproduction spectra at $W=75\gev$ compared with the H1 data~\cite{Aktas:2005xu}. In the dashed lines only the color charge fluctuations contribute to the fluctuations, and in solid lines additional geometry and density fluctuations are included with the parameters constrained by this data. Figure based on~\cite{Mantysaari:2016jaz}. }  

			\label{fig:spectra_cgc}
\end{figure}

The color charge fluctuations result in a finite incoherent cross section even if no geometry fluctuations are introduced. However, this contribution is small (suppressed by $1/N_c^2$)~\cite{Dominguez:2008aa,Marquet:2009vs} and in practice results in an incoherent cross section significantly underestimating the HERA data on $J/\psi$ photoproduction as shown by dashed lines in Fig.~\ref{fig:spectra_cgc}.

\begin{figure}[tb]
		\includegraphics[width=\columnwidth]{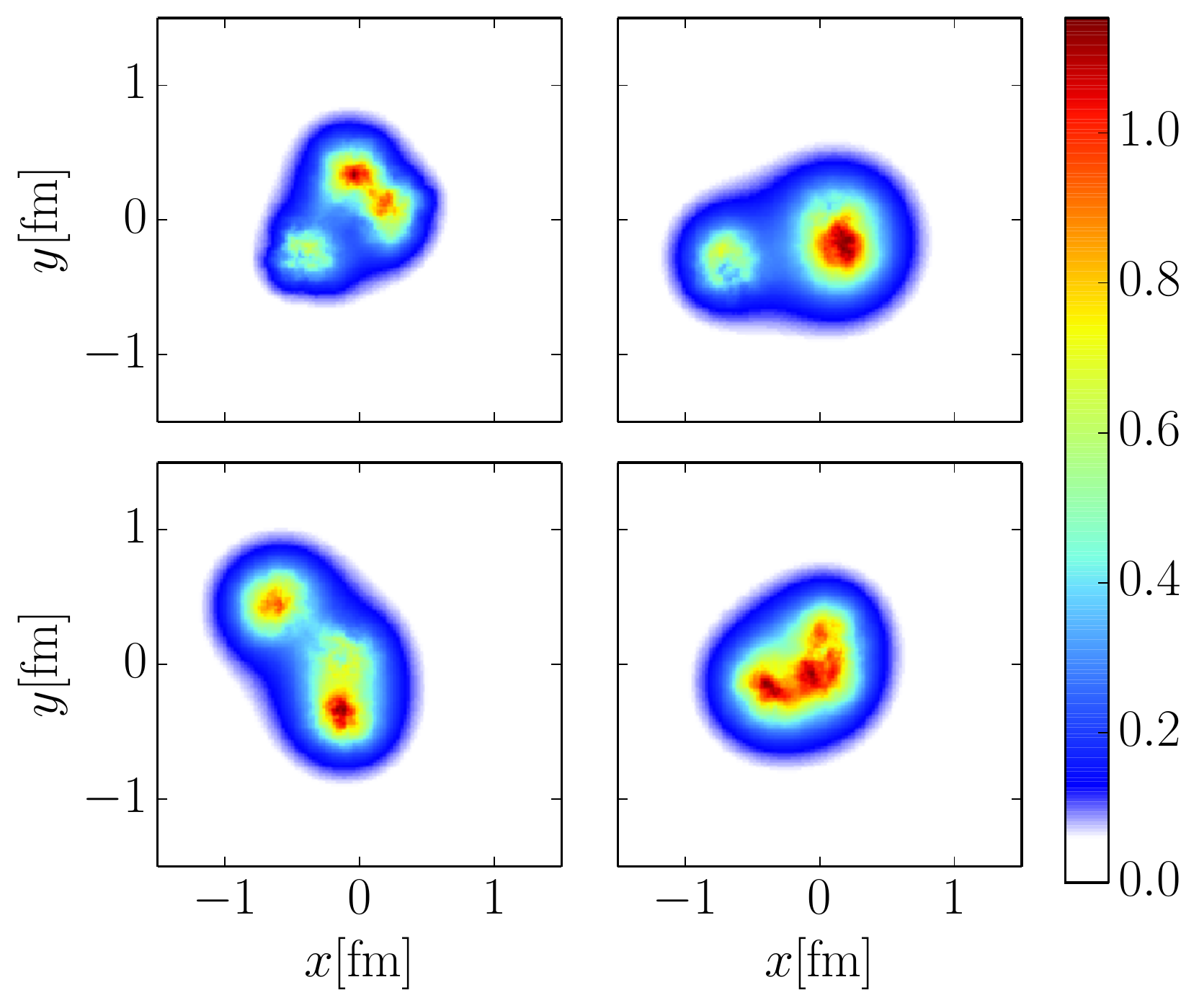} 
				\caption{Example proton density profiles (illustrated by the trace of the Wilson lines, plotting $1 − \mathrm{Re} \tr V (x,y)/N_c$), when the geometry is fixed by the HERA $J/\psi$ production data at $\xpom \approx 10^{-3}$. Figure from Ref.~\cite{Mantysaari:2016jaz}.  }  

			\label{fig:density_cgc}
\end{figure}

In Ref.~\cite{Mantysaari:2016jaz}, geometry and density fluctuations were added to the CGC framework described above. Using the hot spot geometry with three hot spots and fitting the hot spot size and their average separation it is possible to obtain an excellent description of the H1 photoproduction spectra as shown in Fig.~\ref{fig:spectra_cgc} (solid lines). The resulting proton fluctuations (at $\xpom\approx 10^{-3}$ corresponding to the kinematics of the H1 data) are illustrated in Fig.~\ref{fig:density_cgc}, where four examples of the sampled protons are shown. The proton density is illustrated by showing $1 - \Re \mathrm{Tr} V(\xt)/\nc$. Note that as individual Wilson lines are not gauge invariant (unlike $\tr V^\dagger V$), this should be  considered only as an illustration in  arbitrary units. However, the conclusion is similar as in the simpler framework where the IPsat parametrization is used: large event-by-event fluctuations are needed to describe the incoherent vector meson data from HERA, even when the color charge fluctuations are included. Additionally, it is shown that by just introducing large overall density fluctuations it is not possible to obtain a correct incoherent spectrum~\cite{Mantysaari:2016jaz}.

This relatively simple geometry has been further developed in Ref.~\cite{Traini:2018hxd} using an explicit quark model calculation for the wave function of the valence quarks, where correlations between the constituent quarks have been taken into account. The spin dependence of the forces between the quarks results in both attractive and repulsive interactions. All parameters except the size of the small-$x$ gluon cloud around the valence quark can be fixed by the low energy proton and neutron structure data, and consequently only one parameter (size of the gluon cloud) needs to be adjusted to simultaneously describe the coherent and incoherent vector meson production data from HERA. The proton shape fluctuations extracted in Ref.~\cite{Traini:2018hxd} are very similar to the results summarized above.

\subsection{Energy dependence of the fluctuating structure}
\label{sec:edep}
In the discussion above the proton shape fluctuations were constrained at fixed Bjorken-$x$, without considering the $x$ dependence of the proton structure. On the other hand, there are clear measurements that show that the transverse size of the proton grows with decreasing $x$. For example, when the $t$ slope  $B_p$ of the coherent photoproduction cross section, see Eq.~\eqref{eq:coh_tslope}, is extracted at different center-of-mass energies $W$ (corresponding to different $x$), a clear logarithmic dependence on the energy is observed:
\begin{equation}
\label{eq:b_wdep}
 B_p \approx B_0 + 4 \alpha' \ln \frac{W}{90 \gev}.
\end{equation} 
The parameters measured by H1~\cite{Aktas:2005xu} are $B_0=4.630 \pm 0.060 ^{+0.043}_{-0.163} \gev^{-2}$ and $\alpha'=0.164\pm 0.028\pm 0.030 \gev^{-2}$.

In addition to  the growing proton size, fluctuations should evolve in energy. At asymptotically large energies, where the black disc limit is excepted to be reached e.g. as a result of the small-$x$ evolution described in the Color Glass Condensate framework, the probability for the dipole to scatter saturates to unity, and there are no event-by-event fluctuations. In this limit, the incoherent cross section is expected to be suppressed, as event-by-event fluctuations are only possible at the dilute edge of the proton. The energy dependence of the coherent and incoherent cross section measured at HERA~\cite{Alexa:2013xxa} show that the incoherent-to-coherent cross section ratio decreases with increasing center-of-mass energy. Recently, the ALICE collaboration has studied $J/\psi$ photoproduction at different photon-proton center-of-mass energies~\cite{TheALICE:2014dwa,Acharya:2018jua} in ultraperipheral collisions~\cite{Bertulani:2005ru,Klein:2019qfb}. The energy dependence of the incoherent cross section is not measured so far, but the two muon (decay products of $J/\psi$) transverse momentum distributions are compatible with the incoherent cross section being suppressed compared to the coherent one at high energies.

In phenomenological calculations different approaches have been taken to describe the energy dependence of the fluctuations. In the framework developed in Ref.~\cite{Cepila:2016uku} and further applied in Refs.~\cite{Cepila:2017nef,Cepila:2018zky} the proton is assumed to consist of hot spots, whose number $N_{hs}$ depends on the Bjorken-$x$ as
\begin{equation}
\label{eq:nhs}
N_{hs}=p_0 x^{p_1}(1+ p_2 \sqrt{x}).
\end{equation}
This functional form is motivated by a generally used parametrization of the gluon distribution at the initial scale used in  parton distribution function fits (see also Refs.~\cite{Liou:2016mfr,Domine:2018myf} for a recent discussion of the gluon density fluctuations in dilute hadrons).  The number of gluons is then related to the number of hot spots, and the locations for the hot spots are sampled randomly similarly as described in Sec.~\ref{sec:fluct_fixed_x}. The model parameters $p_0=0.011$, $p_1=-0.58$ and $p_2=250$ have been fixed by comparing to energy dependence of the incoherent $J/\psi$ photoproduction data form H1~\cite{Alexa:2013xxa}. 
In this approach a good description of the energy dependence of both coherent and incoherent H1 data is obtained~\cite{Cepila:2016uku}. 

\begin{figure}[tb]
		\includegraphics[width=\columnwidth]{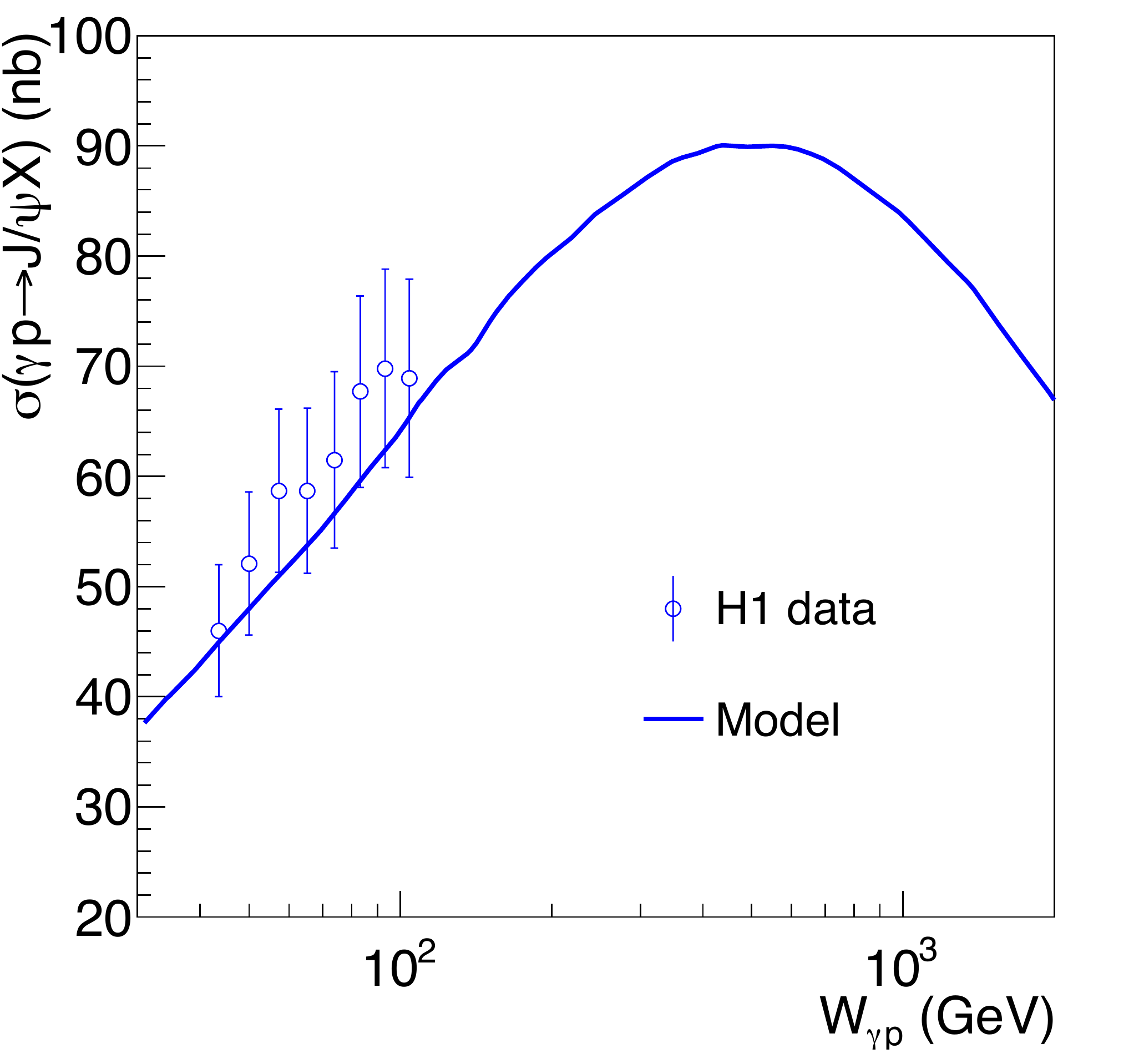} 
				\caption{Energy dependence of the incoherent $J/\psi$ photoproduction cross section from the model where the number of hot spots  in the proton increases with increasing energy. Figure from Ref.~\cite{Cepila:2016uku}.}  

			\label{fig:incoh_wdep_hotspots}
\end{figure}

A genuine prediction of this framework is that as the number of hot spots in a fixed transverse area increases, the event-by-event fluctuations will eventually disappear. This results in an incoherent cross section which has a maximum at the center-of-mass energy $W\sim 700\gev$ (in case of $J/\psi$ photoproduction), after which it decreases with increasing energy, as shown in Fig.~\ref{fig:incoh_wdep_hotspots}. In ultraperipheral proton-lead collisions at the LHC, center of mass energies in the TeV range for the photon-proton system can be achieved. A future measurement of the energy dependence of the cross section ratio will be a crucial test for the model. However, as the proton size evolution is not included in this framework, the model would not reproduce the experimentally observed evolution of the $t$ slope of the coherent cross section shown in Eq.~\eqref{eq:b_wdep}.

A different approach to the energy evolution was taken in Ref.~\cite{Mantysaari:2018zdd}, where the perturbative JIMWLK evolution equation describing the Bjorken-$x$ dependence was applied. There, using the Color Glass Condensate framework described above the proton fluctuating structure at initial Bjorken-$x$ was constrained by comparing with the $J/\psi$ production data at $W=75\gev$.  On top of geometry fluctuations, also overall density fluctuations for each of the three hot spots sampled at initial $x$ are included. Then, the different proton configurations were separately evolved to smaller $x$ by solving the JIMWLK equation, and using the evolved protons the cross sections at higher center-of-mass energies were obtained.

\begin{figure}[tb]
		\includegraphics[width=\columnwidth]{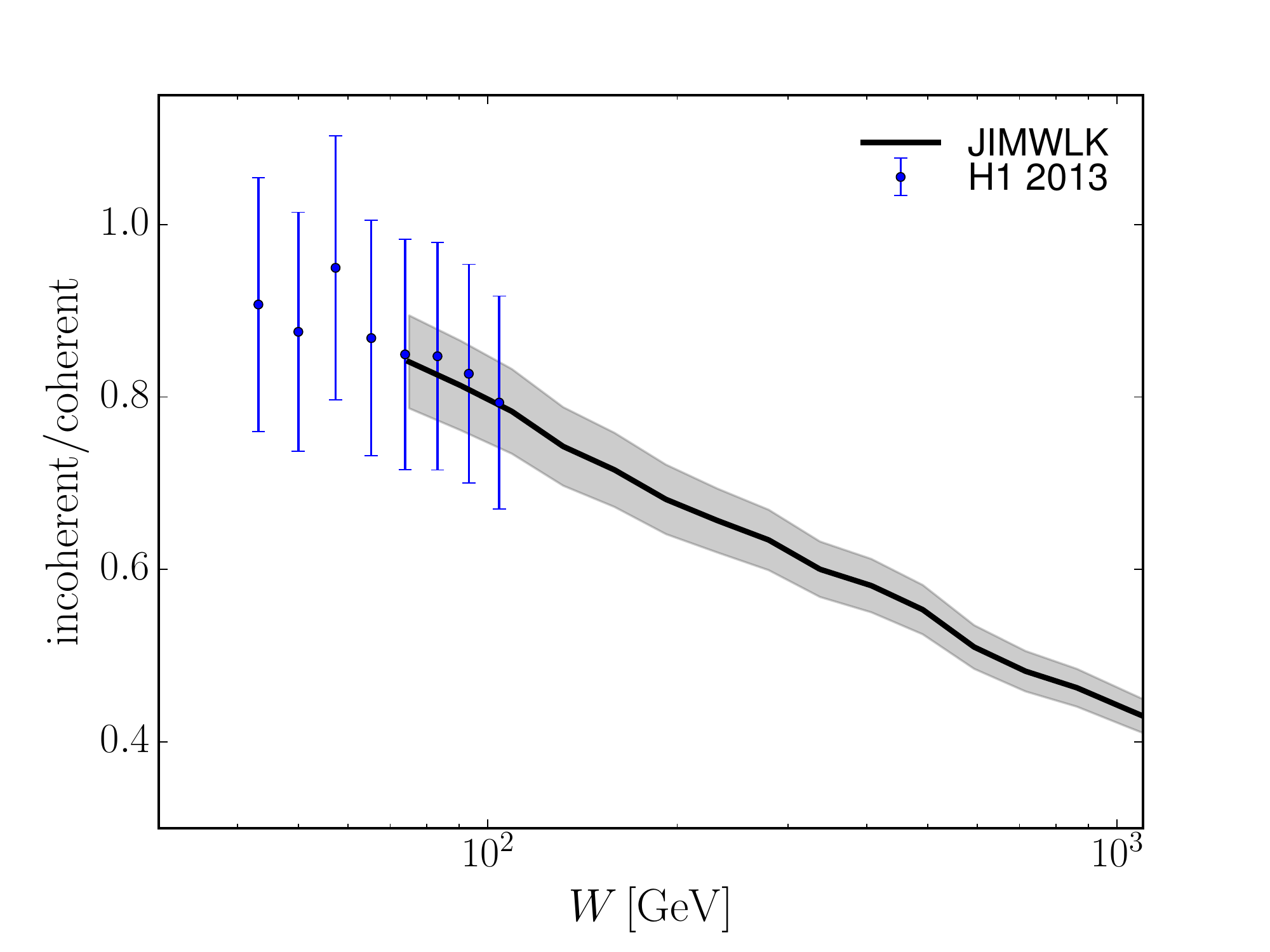} 
				\caption{Incoherent-to-coherent $J/\psi$ photoproduction cross section ratio as measured by H1~\cite{Alexa:2013xxa}, compared with the prediction based on JIMWLK evolution. Figure based on Ref.~\cite{Mantysaari:2018zdd}.}  

			\label{fig:incohcohratio}
\end{figure}
\begin{figure}[tb]
		\includegraphics[width=\columnwidth]{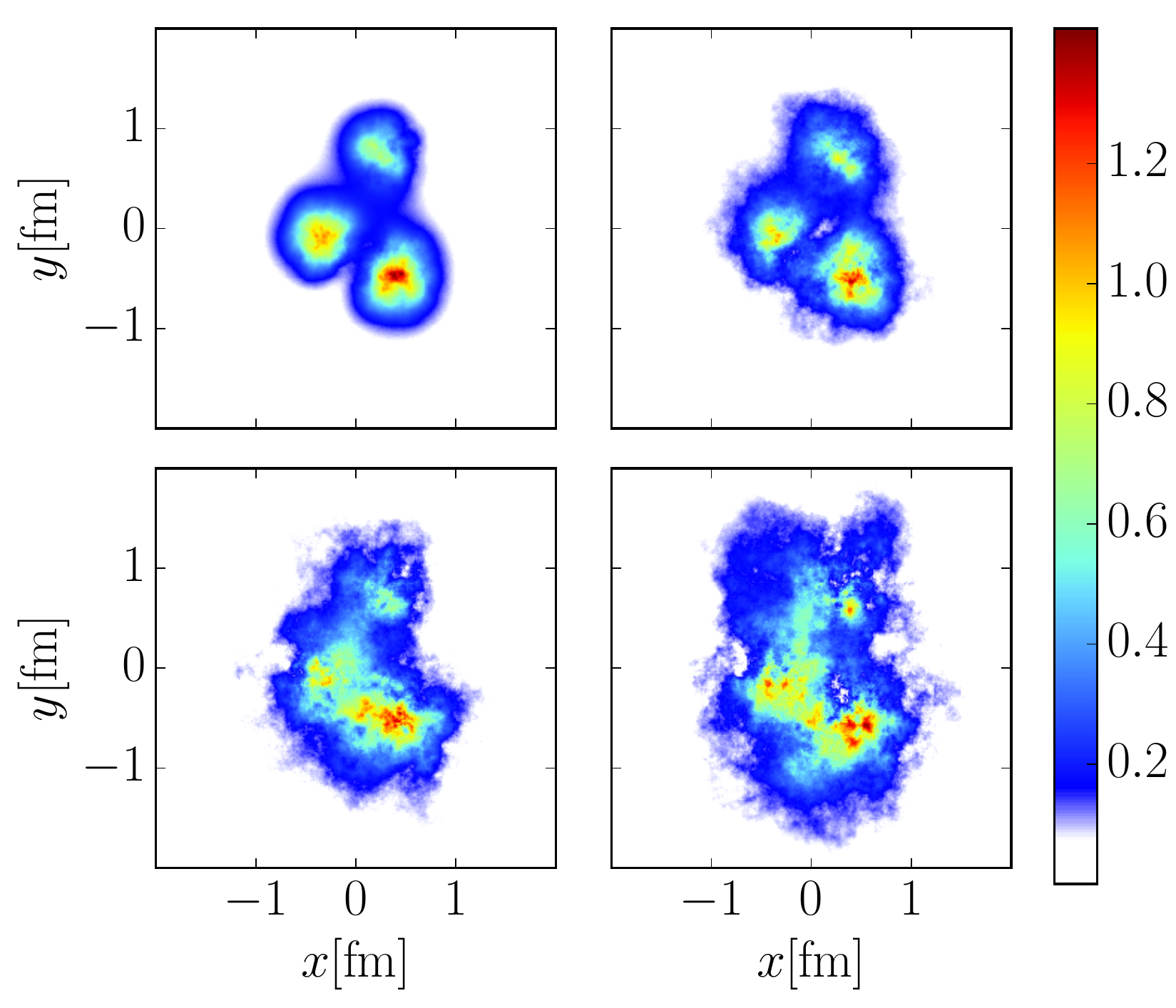} 
				\caption{Example of the proton shape evolution from $x_0=0.01$ to small $x=x_0e^{-y}$ at rapidities $y=0.0, 1.8, 3.5$ and $y=5.3$ from Ref.~\cite{Mantysaari:2018zdd}. The density is illustrated by calculating $1.0 − \mathrm{Re} \tr V (x,y)/N_c$. }  

			\label{fig:proton_shape_evol}
\end{figure}

The calculated incoherent-to-coherent cross section ratio as a function of photon-proton center-of mass energy $W$ is shown in Fig.~\ref{fig:incohcohratio}. The decreasing ratio shows that the evolution washes out the initial hot-spot structure. This can be seen from the actual evolution of a one particular proton configuration  illustrated in Fig.~\ref{fig:proton_shape_evol}. The free parameters in the evolution, the value for the fixed QCD coupling constant $\as$ and the infrared regulator limiting the unphysical growth of the proton size due to the missing confinement scale effects, are fixed by requiring a compatible evolution speed with the inclusive charm structure function data.   However, a disadvantage of this approach is that the simultaneous description of both inclusive and exclusive observables is not possible without introduction of a non-perturbative contribution, due to the numerically large contributions from non-perturbatively large dipoles (see~\cite{Mantysaari:2018nng} for a detailed analysis of large dipole contributions).


\section{Proton shape from proton-proton collisions}
\label{sec:pp}
So far it has been discussed how the proton shape fluctuations can be constrained based on exclusive vector meson production in deep inelastic scattering. Qualitatively similar results have been obtained based on analyses of other observables, and these findings support the picture developed above where the proton shape has large event-by-event fluctuations.

\subsection{Elastic scattering}

In Ref.~\cite{Albacete:2016pmp}  elastic proton-proton scattering at small momentum transfer $|t|$ is studied. This process is measured at $\sqrt{s}=2.76, 7$ and $13\tev$ center-of-mass energy at the LHC~\cite{Antchev:2011zz,Antchev:2018rec,Antchev:2018edk} (only $7 \tev$ data is used in the analysis of Ref.~\cite{Albacete:2016pmp}), and at smaller $\sqrt{s}=62.5\gev$ at ISR~\cite{Amaldi:1979kd}. As there is no hard scale, perturbative calculations can not be applied to describe this process. Instead, the  real and imaginary part of the scattering amplitude can be parametrized, and the values for the unknown parameters can be constrained by fitting the $t$ spectra and requiring that that the real-to-imaginary part ratio satisfies the measured value\footnote{In case of LHC, the authors use an extrapolated value $\rho=0.14^{+0.01}_{-0.08}$ from~\cite{Cudell:2002xe} which is compatible with the more recent TOTEM measurement $\rho=0.09/0.10\pm 0.01$  (depending on physics assumptions)~\cite{Antchev:2017yns}.}.

A striking observation in elastic scattering processes is the so called Hollowness or grayness effect~\cite{Alkin:2014rfa,Dremin:2015ujt,Troshin:2016frs,Arriola:2016bxa}: the inelasticity density describing how different impact parameters contribute to the inelastic cross section  has a maximum at a non-zero impact parameter at LHC energies\footnote{The hollowness effect could be seen to suggest that the black disk limit discussed in Sec.~\ref{sec:edep} it not necessarily reached at asymptotically high energies.}. On the other hand, at lower ISR energies  the maximum is at $\bt=0$ at ISR energies. The inelasticity density is defined in terms of the elastic scattering amplitude $T_\text{el}$ as
\begin{equation}
\label{eq:ineldens}
G_\text{in}(\sqrt{s},\bt) \equiv \frac{\der^2 \sigma_\text{inel}}{\der^2 \bt} = 2\Im T_\text{el}(\sqrt{s},\bt) - |T_\text{el}(\sqrt{s},\bt)|^2.
\end{equation}

Access to the proton geometry is obtained by assuming that the  proton consists of $N_\text{hs}=3$ hot spots, and when the two hot spots collide, the scattering amplitude is assumed to be Gaussian in the transverse distance between the two hot spots. The positions for the hot spots are sampled from a Gaussian distribution, on top of which correlations are included. The distribution for the hot spots is given as
\begin{multline}
\label{eq:corr_density}
D(\{\bt_i\}) = C \left[\prod_{i=1}^{N_\text{hs}} d(\bt_i) \right] \delta^{(2)}(\bt_1+\bt_2+\bt_3) \\
\times \prod_{i,j=1;i<j}^{N_\text{hs}} \left(1 - e^{- |\bt_i - \bt_j|^2/r_c^2} \right)
\end{multline}
where $\bt_i$ are the transverse positions for the hot spots, $d(\bt_i)$ is a Gaussian distribution and $r_c$ parametrizes  the range of short-range repulsive correlations. Note that in the limit $r_c\to 0$ this is similar to the hot spot structure parametrization of Eq.~\eqref{eq:hotpsottp}.

\begin{figure}[tb]
		\includegraphics[width=\columnwidth]{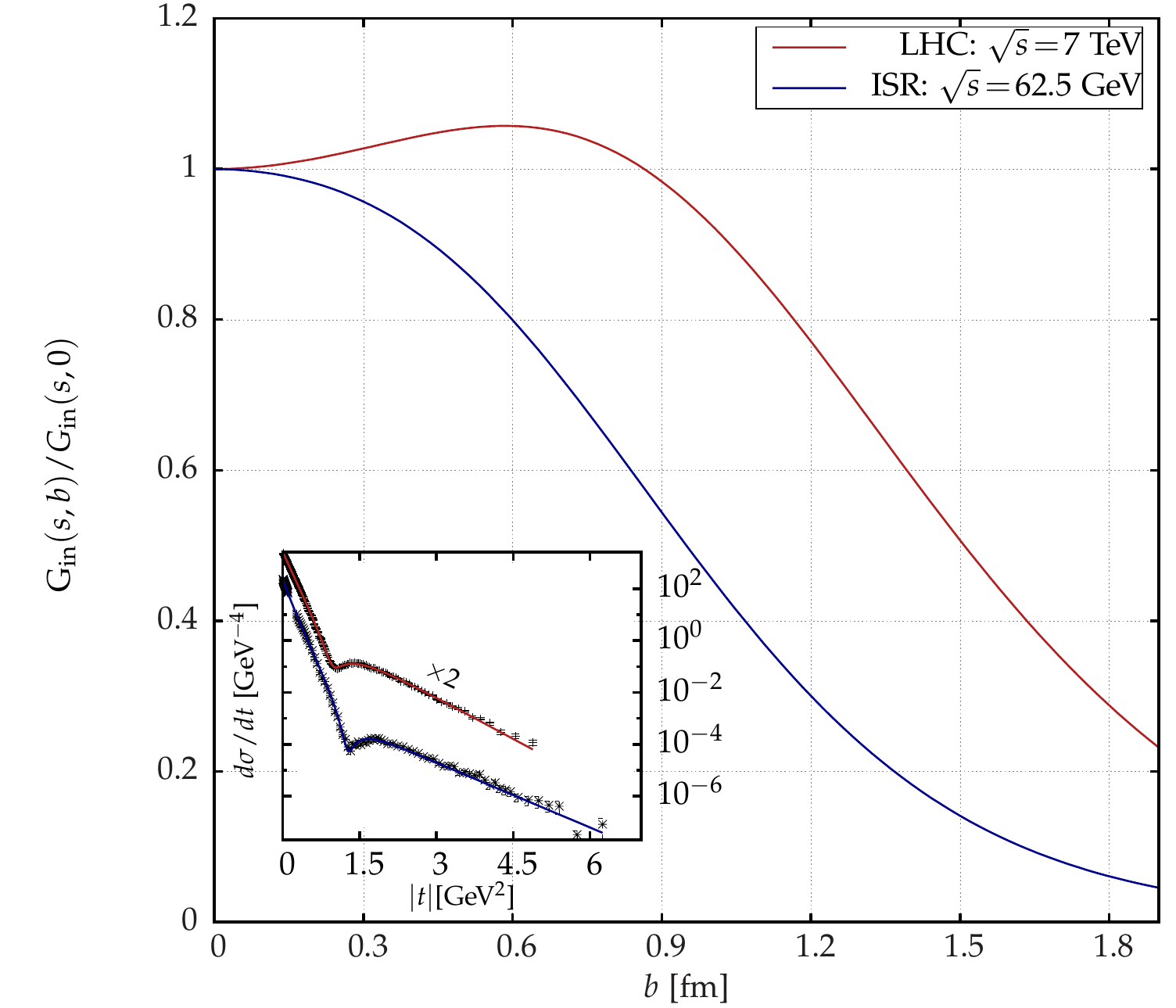} 
				\caption{Inelasticity density at low and high center-of-mass energies. The panel at the bottom shows the $t$ spectra from which the density profile is extracted. Figure from Ref.~\cite{Albacete:2016pmp}. }  

			\label{fig:inel_density}
\end{figure}

In Ref.~\cite{Albacete:2016pmp} it is shown that it is only possible to reproduce the Hollowness effect in this picture if there are repulsive short range correlations between the hot spots.  The extracted inelasticity densities at ISR and LHC energies are shown in Fig.~\ref{fig:inel_density}, using the proton geometry with $r_c=0.3\fm$ parameter to set the scale for repulsive interactions. Compared to the exclusive $J/\psi$ photoproduction discussed above, the proton size is much larger in Ref.~\cite{Albacete:2016pmp} which is obtained mainly by having a much longer distance between the hot spots. A larger proton is required, as the elastic proton-proton spectrum is much more steep than the coherent $J/\psi$ spectrum.The parameters can not be directly compared, however. The $J/\psi$ photoproduction probes the gluon density at perturbative scales, whereas elastic proton-proton scattering is a non-perturbative process in the applied kinematics. Additionally, these processes are sensitive to the proton structure at different Bjorken-$x$.
However, the hot spots are constrained in Ref.~\cite{Albacete:2016pmp} to be much smaller than the proton size which is qualitatively compatible with the large event-by-event fluctuations constrained by studying the $J/\psi$ photoproduction.

\subsection{Symmetric cumulants}
The same proton geometry that is constrained by elastic proton-proton data discussed above was applied in studies of symmetric flow cumulants in a hydrodynamical model in Ref.~\cite{Albacete:2017ajt}. The (normalized) symmetric cumulant $\text{NSC}(n,m)$ is defined as
\begin{equation}
\text{NSC}(n,m) = \frac{\langle v_n^2 v_m^2 \rangle - \langle v_n^2\rangle \langle v_m^2 \rangle }{\langle v_n^2\rangle \langle v_m^2\rangle },
\end{equation}
and it measures how the flow harmonics  $v_n$ characterizing collective features of the event are correlated. In the absence of correlations, NSC$(n,m)=0$. The flow harmonics $v_n$ are obtained when the azimuthal angle $\phi$ distribution of particles is written as a Fourier decomposition
\begin{equation}
\label{eq:vn}
\frac{\der N}{\der \phi} = v_0(1 + 2 v_2 \cos [2(\phi - \Phi_2)] + 3 v_3 \cos[3(\phi - \Phi_3)] \dots).
\end{equation} 
Here $\Phi_n$ refers to the event plane angle, defined as the first angle where the $n$th harmonic has its maximum value. In heavy ion collisions, the initial state spatial anisotropy can be  translated into momentum space anisotropies (quantified by the $v_n$ coefficients) as a result of the hydrodynamical evolution of the produced Quark Gluon Plasma. 

The approach in Ref.~\cite{Albacete:2017ajt} is to calculate the initial state spatial eccentricities and relate those to the final state momentum space anisotropies. To determine the eccentricities, one first samples which hot spots collide. The collision probability is obtained from the inelasticity density,  Eq.~\eqref{eq:ineldens}, with the scattering amplitude $\sim e^{-d^2/R_{hs}^2}$ where $R_{hs}$ is the hot spot size constrained by the elastic proton-proton scattering data. Subsequently, each collided hot spot deposits a random amount of entropy around it. The eccentricities are then calculated from the total entropy deposition profile. These initial state geometric anisotropies are then translated into momentum space by assuming that the hydrodynamical response to the initial state anisotropies is linear. In case of lowest harmonic coefficients $v_2$ and $v_3$, this is a good approximation  in high multiplicitly  nuclear collisions~\cite{Niemi:2012aj}, and an approximatively linear relation is found also in hydrodynamical simulations of proton-lead collisions~\cite{Bozek:2013uha} (see also Ref.~\cite{Schenke:2019pmk})  .  
\begin{figure}[tb]
		\includegraphics[width=\columnwidth]{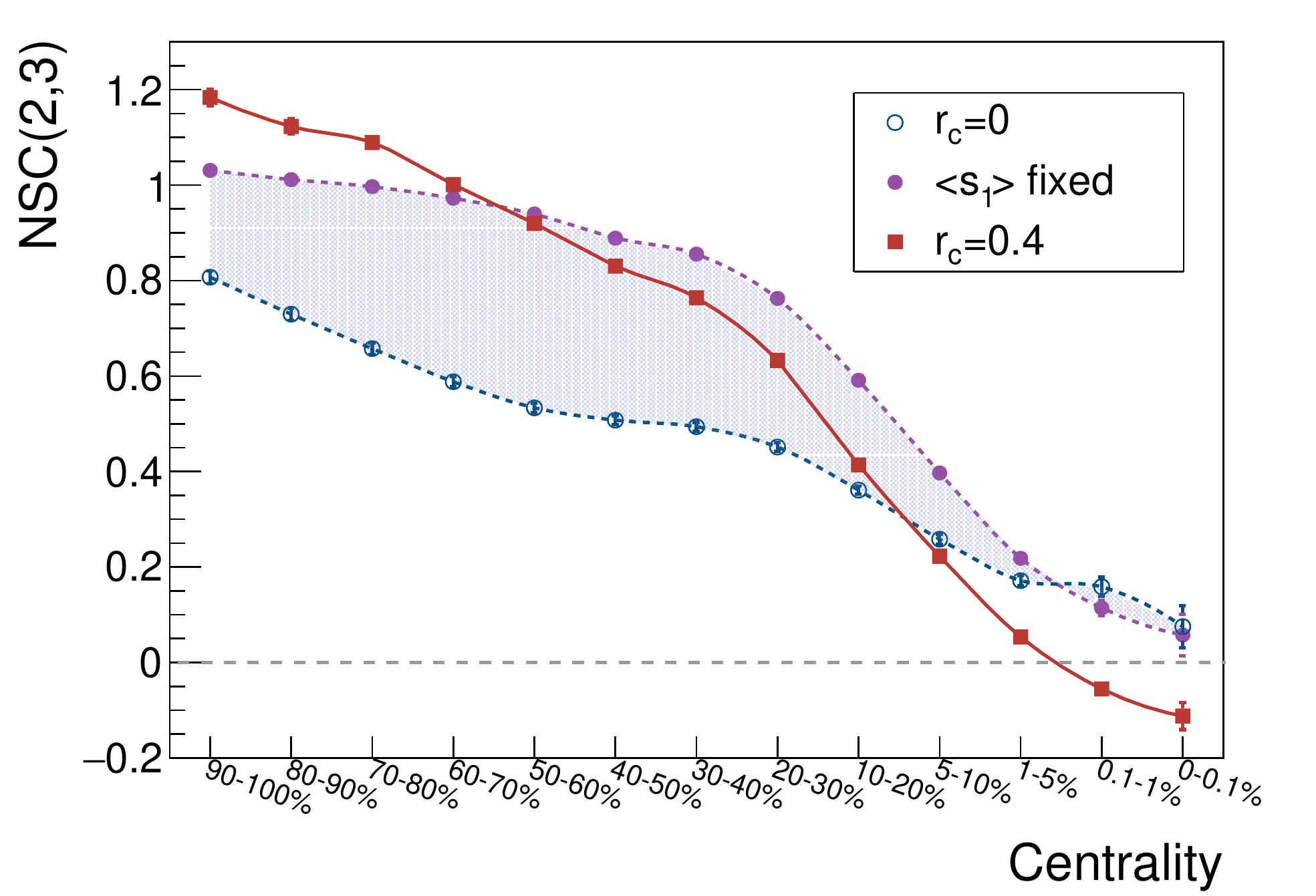} 
				\caption{Normalized symmetric cumulant in proton-proton collisions as a function of centrality. The $r_c=0.4$ calculation includes repulsive short range correlations not included in the results labeled as $r_c=0$ and $\langle s_1\rangle$. In the latter case  the distribution in Eq.~\eqref{eq:corr_density} is adjusted to result in the same root mean square radius for the proton as in the case with repulsive correlations. Figure from Ref.~\cite{Albacete:2017ajt}.}  

			\label{fig:nsc}
\end{figure}

The ALICE and CMS collaborations have observed that the correlation between $v_2$ and $v_3$ becomes negative in  proton-proton collisions in the highest multiplicity bins~\cite{Sirunyan:2017uyl,Acharya:2019vdf}.  In Ref.~\cite{Albacete:2017ajt}, the total entropy production is taken to serve as a proxy for the experimental multiplicity (or centrality) classes. 
The calculated symmetric cumulants NSC$(2,3)$ are shown in Fig.~\ref{fig:nsc}. The calculation $r_c=0$ corresponds to the case where there are no repulsive short range correlations in the proton structure. As a result, the correlation between the second and third order cumulants is always positive. Instead, when a relatively large short range repulsive correlation $r_c=0.4\fm$ is included, the correlation becomes negative in most central collisions, which is qualitatively compatible with the CMS data. Note that the same repulsive correlation was required to describe the Hollowness effect as discussed above. Including the full hydrodynamical evolution in the framework is  underway~\cite{Albacete:2018rzf}.

\section{Fluctuating nucleons in nuclei}
\label{sec:nuclei}
\subsection{Vector meson production}
So far the in this Review the focus has been on the role of proton shape fluctuations on exclusive, elastic and collective observables. However, there is nothing special about the proton target, and  similar analyses can be performed with both light and heavy nuclei. In nuclei, fluctuations are excepted to take place at multiple different distance scales. At longest distances, the positions of the nucleons fluctuate, following the Woods-Saxon distribution in heavy nuclei.  At shorter scales, the nucleon shape fluctuations as discussed above can become visible. Additionally, there are color charge fluctuations in the hadrons, and the color charges are correlated over the distance scale $1/Q_s$, where $Q_s$ is the saturation scale. At high energies or in heavy nuclei where the saturation scale is large (we have an approximate scaling $Q_s^2 \sim A^{1/3} x^{-0.2}$), the color charge fluctuations manifest themselves at distance scale much smaller than the hot spot size.

\begin{figure}[tb]
		\includegraphics[width=\columnwidth]{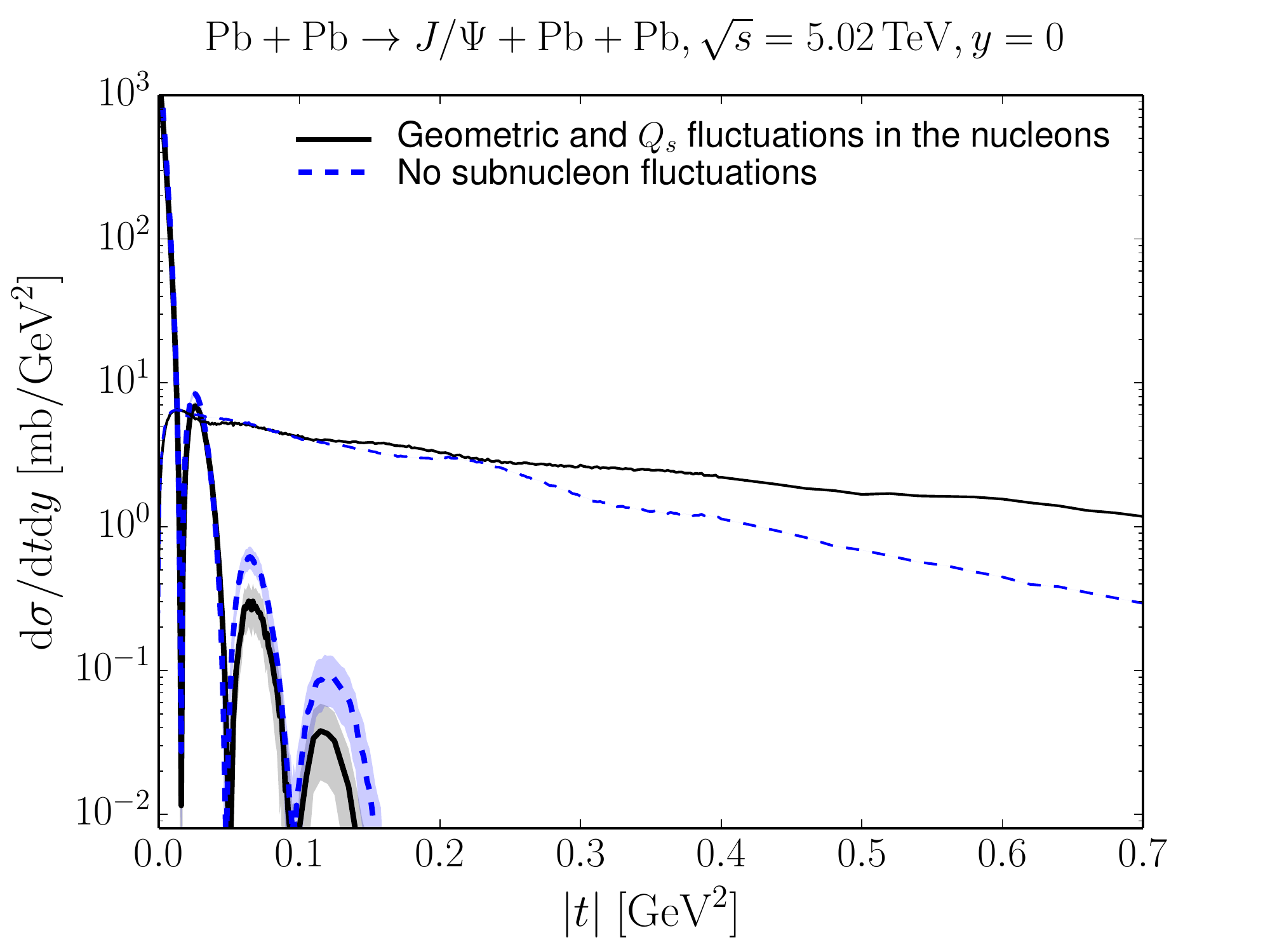} 
				\caption{Predictions for the coherent (thick lines) and incoherent (thin lines) $J/\Psi$ photoproduction in ultraperipheral Pb$+$Pb collisions at the LHC. Solid curves correspond to the case where nucleons consist of fluctuating hot spots, and dashed curves the case where nucleons have no substructure. Figure from Ref.~\cite{Mantysaari:2017dwh}.}  

			\label{fig:pb_t}
\end{figure}

Taking into account the nucleon shape fluctuations, exclusive vector meson production off light nuclei (deteuron and helium) has been studied in Ref.~\cite{Mantysaari:2019jhh}. Heavy nuclei, gold and lead, were first considered in Ref.~\cite{Mantysaari:2017dwh} by extending the calculations of Refs.~\cite{Caldwell:2009ke,Lappi:2010dd,Toll:2012mb,Lappi:2013am} to include the nucleon shape fluctuations on top of the fluctuating nucleon positions from the Woods-Saxon distribution. Additional contribution to the incoherent cross section from the fluctuating nucleon shape was found to improve the simultaneous description of both the coherent and incoherent cross sections measured in ultraperipheral collisions at the LHC. Later the framework where the number of hot spots depends on Bjorken-$x$ was also generalized to this setting~\cite{Cepila:2017nef,Cepila:2018zky}. 

In exclusive vector meson production the distance scale is set by the momentum transfer $t$. As shown in Ref.~\cite{Lappi:2010dd}, if the incoherent cross section is parametrized as $e^{-B|t|}$, then the slope $B$ is set by the size of the object that is fluctuating. As discussed above, there are (at least) two different distance scales at which fluctuations take place: the nucleon size, and the substructure size. In Fig.~\ref{fig:pb_t} the predicted $J/\Psi$ photoproduction cross section is shown as a function of $t$, with  (black solid lines) and without (blue dashed lines) nucleon substructure. 

The coherent cross section is found to be similar with and without nuclear substructure fluctuations as expected, as the average shape of the nucleus does not change significantly when the nucleon substructure fluctuations are included, see discussion in Sec.~\ref{sec:fluct_fixed_x}. The fact that the average density profiles are not exactly identical result in some deviations in the coherent spectra at larger $|t|$. 
On the other hand, in  incoherent scattering the $t$ slopes are clearly different at  $|t|\gtrsim 0.25\gev^2$, which corresponds to the distance scale $\sim 0.4\fm$, which is the scale of the hot spots (see Fig.~\ref{fig:density_cgc}). 

\begin{figure}[tb]
		\includegraphics[width=\columnwidth]{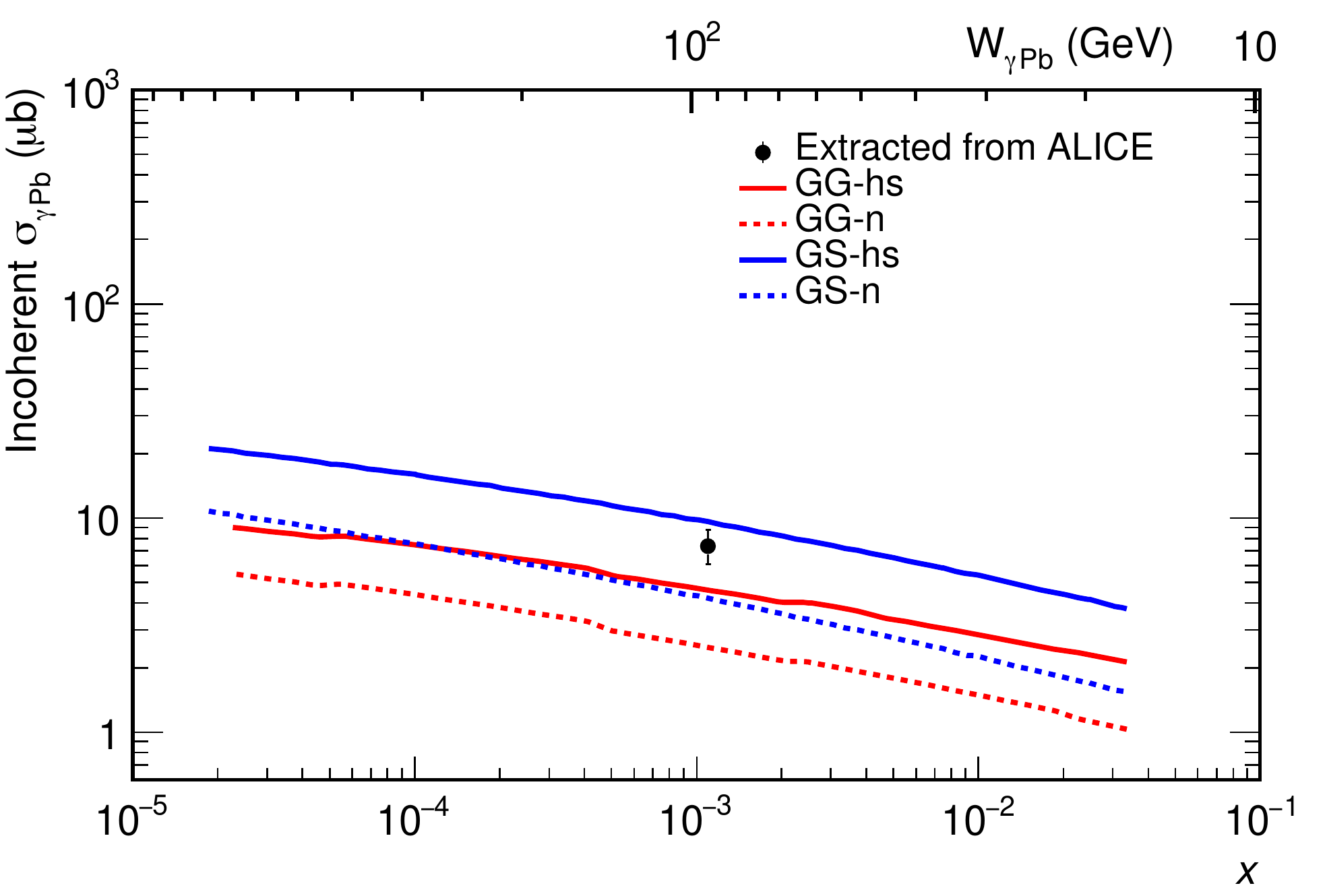} 
				\caption{
			Bjorken-$x$ dependence of the incoherent $J/\Psi$ photoproduction. Predictions shown as solid lines (label \emph{hs}) include nucleons consisting of $x$ dependent number of hot spots, and dashed lines have no nucleon shape fluctuations. Different colors refer to different models for the dipole-proton interaction. Figure from Ref.~\cite{Cepila:2017nef}.}
			\label{fig:pb_y}
\end{figure}

The LHC experiments have only measured the $t$ integrated cross sections, as the experimental methods are based on template fits. For the incoherent $J/\Psi$ photoproduction, there exists only one datapoint from ALICE at $\sqrt{s}=2760\gev$ at midrapidity~\cite{Abbas:2013oua}. At midrapidity there is no ambiquity in determining the Bjorken-$x$, and the $J/\Psi$ photoproduction cross section at $x\approx 10^{-3}$ can be extracted as the photon flux is known. In Fig.~\ref{fig:pb_y} the predicted incoherent $J/\Psi$ photoproduction cross section as a function of $x$ is shown, based on Ref.~\cite{Cepila:2017nef}. The predictions from the calculations that both include and do not include nucleon substructure fluctuations (where the number of hot spots depends on $x$, see Eq.~\eqref{eq:nhs}) are shown. The substructure is shown to enhance the incoherent cross section almost by a factor of $2$, which is consistent with the ALICE data. It is important to note that the predicted absolute normalization depends strongly on the applied $J/\Psi$ wave function, and simultaneous comparison to both coherent and incoherent data is necessary. Cross section ratios are theoretically more robust. In Ref.~\cite{Mantysaari:2017dwh} it was shown that the wave function uncertainty almost cancels in the incoherent-to-coherent ratio, and the nucleon substructure fluctuations improve the description of the ALICE data, which still has a quite large uncertainty (see also Ref.~\cite{Sambasivam:2019gdd}).

\subsection{ Nucleon substructure and flow in heavy ion collisions}
Collective phenomena are traditionally seen as a proof of the formation of the Quark Gluon Plasma (QGP) in heavy ion collisions at high energy. However, recently similar signals have been also observed in high multiplicity  proton-proton and proton-nucleus collisions. This suggests that a hydrodynamically evolving medium may be produced also in smaller collision systems. As the proton shape fluctuations result in initial state anisotropies~\cite{Welsh:2016siu}, the event-by-event fluctuating proton structure can in principle be constrained by comparing hydrodynamical simulations including proton shape fluctuations with  experimental data. 
However, there are additional sources of momentum correlations that do not require a medium formation and could contribute significantly to the measured collective effects. 
Also, not all effects traditionally associated with the QGP production are seen in small system collisions. For example, production of high transverse momentum particles is suppressed  in heavy ion collisions, and this effect is understood to be caused by partons losing energy when propagating through the medium. Similar effects are not seen in even the highest multiplicity proton-proton and proton-nucleus collisions. For a review, the reader is referred to Ref.~\cite{Dusling:2015gta}.

Theoretically it is also not a priori clear if a locally thermalised fluid can be produced in such a small systems. The applicability of hydrodynamics can also be limited by the presence of large density gradients~\cite{Niemi:2014wta}, at least if the multiplicities are not very high. The highest multiplicity events, on the other hand, may only be produced by a subset of rare proton configurations~\cite{Coleman-Smith:2013rla}.

In Sec.~\ref{sec:fluct_fixed_x} it was shown how the Wilson lines, encoding all the information about the small-$x$ structure, can be constrained based on diffractive vector meson production data. This can be taken as an input for hydrodynamical simulations of the proton-nucleus (and in principle proton-proton) collisions e.g. in the 
framework developed in Ref.~\cite{Schenke:2012wb}. There, the developed IP-Glasma model is first used to determine the initial state after the collision, and to solve the early time evolution (see also Refs.~\cite{Giacalone:2019kgg,Giacalone:2019vwh} for a recent develompent of the fluctuating initial condition for the hydrodynamical evolution from the CGC framework).  When the system is close to the local thermal equilibrium, hydrodynamical evolution is applied using  the relativistic viscous hydrodynamic simulation MUSIC~\cite{Schenke:2010nt,Schenke:2010rr}. After the hydrodynamical phase the final final stages of the evolution for the produced particles are simulated by using the UrQMD model which implements the  microscopic Boltzmann equation~\cite{Bass:1998ca,Bleicher:1999xi}.

In general, as the hydrodynamical evolution translates spatial anisotropies into the momentum space correlations, if the colliding proton is spherical, only small momentum space anisotropies can be expected. In the IP-Glasma+hydrodynamics framework this expectation was confirmed in Ref.~\cite{Schenke:2014zha}.  When the proton shape fluctuations determined in Sec.~\ref{sec:fluct_fixed_x} were used, a good description of the measured elliptic and triangular flow coefficient $v_2$ and $v_3$ defined in Eq.~\eqref{eq:vn} in high-multiplicity proton-lead collisions were obtained \cite{Mantysaari:2017cni}. 
Similarly, in Ref.~\cite{Weller:2017tsr} it was found that a three valence quark structure for the nucleons is necessary in hydrodynamical simulations of proton-proton, proton-lead and lead-lead collisions. For the effect of proton substructure on the initial state structure (e.g. eccentricities), see Ref.~\cite{Welsh:2016siu}.

Another approach to determine the proton shape fluctuations is to first consider hydrodynamical simulations of  lead-lead and proton-lead collisions with different nucleon shape fluctuations, and by comparing with experimental measurements to deduce the preferred shape fluctuations. A major complication in this approach is that the hydrodynamical simulations are computationally intensive, and it is in practice not possible to produce enough statistics with different descriptions of the fluctuating initial state to perform a straightforward minimization procedure.

A powerful approach that can be used to determine the initial state  is based on Bayesian statistics. In the context of heavy ion collisions, this approach has been used  recently to constrain the fundamental properties of the produced Quark Gluon Plasma (QGP). These properties include, for example, the temperature dependence of the shear and bulk viscosities and the temperature at which the hydrodynamical evolution ends and the medium is converted into particles whose subsequent evolution is also included by using the UrQMD model. The method developed in Ref.~\cite{Bernhard:2016tnd} (see recent highlights in Ref.~\cite{Bernhard:2019bmu}) uses a Gaussian process emulator that is trained to effectively interpolate in the multi dimensional parameter space. Applying Bayesian statistics and numerically light emulator techniques, it becomes possible to construct the likelihood distributions for the model parameters. As a result, fundamental properties of the QGP with uncertainty estimates are obtained. 

To determine the proton shape fluctuations based on high-multiplicity proton-nucleus data from RHIC and from the LHC, this approach was generalized in Ref.~\cite{Moreland:2018gsh} to include a hot spot structure for the nucleons. The number of hot spots, as well as their size, are model parameters along with the QGP properties mentioned above. Important observables sensitive to the hot spot structure that are used to calibrate the Gaussian emulator are e.g. mean transverse momenta and flow cumulant measurements made in proton-lead collisions. Also, data from lead-lead collisions is included in the analysis.

\begin{figure}[tb]
		\includegraphics[width=\columnwidth]{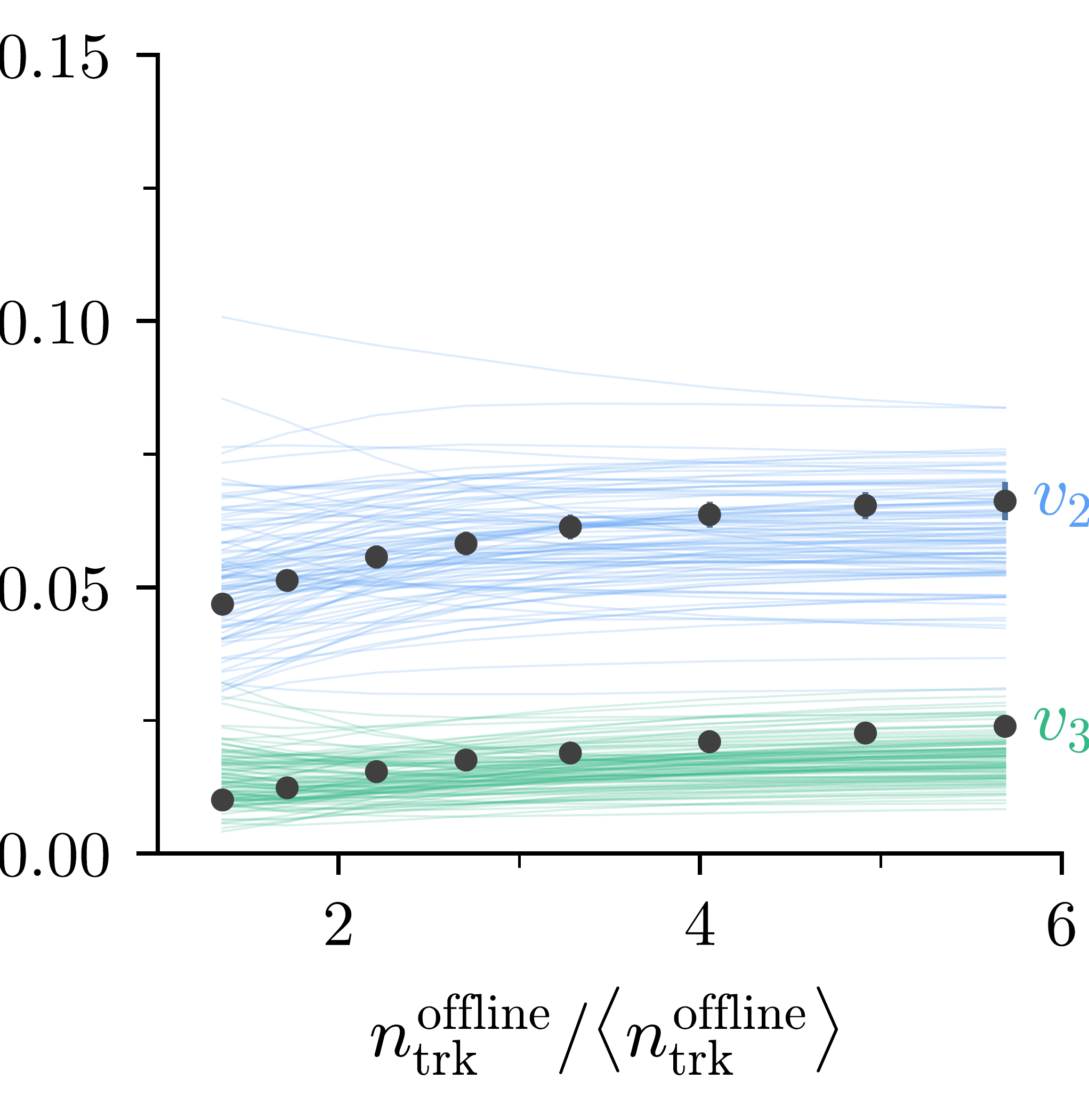} 
				\caption{
			Elliptic and triangular flow $v_n$ as a function of centrality in $p+Pb$ collisions at $\sqrt{s_{NN}}=5.02 \tev$ compared with the CMS data~\cite{Chatrchyan:2013nka}. Different curves correspond to different parametrizations sampled from the posterior distributions. Figure from Ref.~\cite{Moreland:2018gsh}.}
			\label{fig:v2v3}
\end{figure}

From the obtained posterior distributions, it is possible to sample the most likely parametrizations describing the nucleon structure fluctuations, as well as QGP properties and the correlations between the different parameters. In Fig.~\ref{fig:v2v3} the centrality dependence of the elliptic ($v_2$) and triangular ($v_3$) flow coefficients are calculated from 100 random parametrizations sampled from the posterior distribution, and compared with the CMS data~\cite{Chatrchyan:2013nka} (also used to train the emulator). A reasonable description of the flow harmonics, as well as of the centrality dependence of the charge particle yield, are obtained. Simultaneously, e.g. the flow harmonics measured in lead-lead collisions are described accurately and with a much smaller variation between the different parametrizations sampled from the posterior distribution than in the case of proton-lead collisions. 

\begin{figure}[tb]
		\includegraphics[width=0.9\columnwidth]{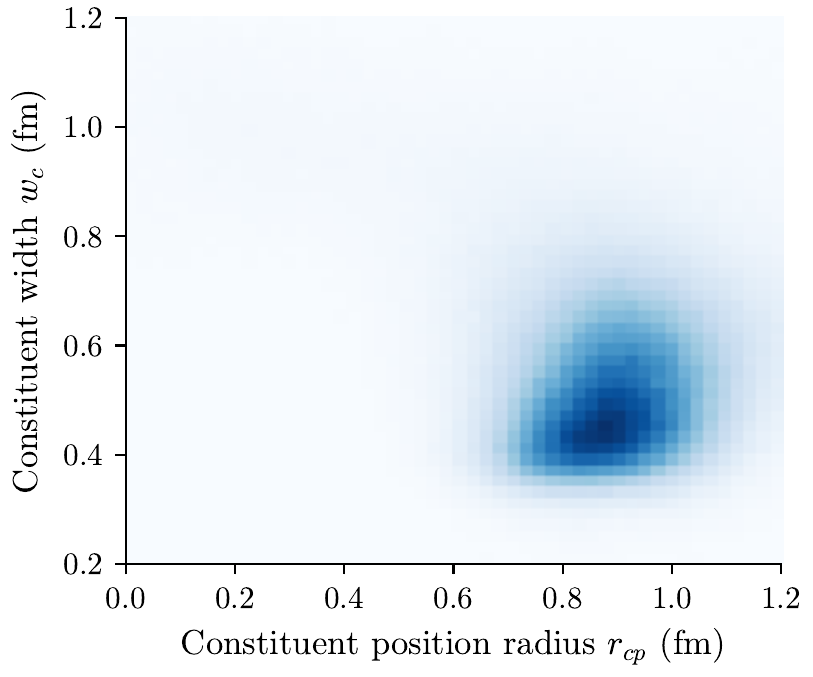} 
				\caption{
			Probability distribution for the hot spot (\emph{constituent}) width, and the proton width (parametrized as \emph{Constituent position radius}, based on the Bayesian analysis of proton-lead and lead-lead data. Figure from Ref.~\cite{Moreland:2018gsh}.}
			\label{fig:bayesianproton}
\end{figure}

For the nucleon, the analysis in Ref.~\cite{Moreland:2018gsh} prefers many hot spots ($\gtrsim 5$). For the hot spot size, and for the size of the nucleon, the posterior distribution is illustrated in Fig.~\ref{fig:bayesianproton}. When comparing these results with the proton shapes constrained by diffractive $J/\psi$ production data and succesfully used to describe proton-lead collisions in the IP-Glasma$+$hydrodynamics framework, one notices that the obtained protons and hot spot sizes seem to be much larger. However, the density profiles can not be directly compared. The analysis of Ref.~\cite{Moreland:2018gsh} prefers the entropy production to be proportional to the geometric mean of the density profiles of the two nuclei. Thus, in proton-nucleus collisions the initial entropy deposition scales as $\der S(\bt) \sim \sqrt{T_A(\bt) T_p(\bt)}$, where $T_A(\bt)$  and $T_p(\bt)$ are the transverse density profiles for the nucleus and for the proton, respectively. On the other hand, in the IP-Glasma framework the initial entropy deposition scales as $\sim T_A(\bt) T_p(\bt)$.

In central (high multiplicity) collisions the nuclear density is approximatively constant and dependence on $T_A(\bt)$ is weak. Consequently, one should compare the squared proton density profile from  Ref.~\cite{Moreland:2018gsh} to the IP-Glasma density profile. For example, the Gaussian width for the hot spot size obtained in Ref.~\cite{Moreland:2018gsh} is $v=0.47^{+0.20}_{-0.15}\fm$ (when the hot spot profile is $ e^{-b^2/(2 v^2)}$). The squared density profile then corresponds to a Gaussian distribution that has a width parameter $v/\sqrt{2}=0.33^{+0.14}_{-0.11}\fm$. This should be compared to the hot spot width $\sqrt{B_q}=0.165\fm$ obtained by fitting the $J/\psi$ data as discussed in Sec.~\ref{sec:fluct_fixed_x}. Although the parametrizations are still not exactly compatible, qualitatively they suggest a similar fluctuating substructure for the proton. On the other hand, the differences demonstrate the limitations of the current approaches to quantitatively extract the fluctuating nucleon shapes from hadronic collisions, as the results depend on the details of the particle production mechanism.


\section{Outlook}

Multiple measurements from different collider experiments suggest that there are significant quantum mechanical event-by-event fluctuations in the transverse profile of the proton at small Bjorken-$x$. This picture is obtained by analysing elastic and diffractive scattering processes, as well as correlations observed in high multiplicity proton-proton and proton-lead collisions. 

Even though the qualitative picture pointing towards sizeable event-by-event fluctuations is the same based on all these observables,  different processes that probe the proton structure at different scales result in e.g. different sizes for the hot spots. At this point it is not clear how, for example, results from elastic low momentum transfer proton-proton collisions can be related to the proton shape fluctuations extracted by studying exclusive production of heavy vector mesons which is a perturbative process. Similarly, in analyses of collective phenomena in high multiplicity collisions one is actually sensitive to the initial energy density or entropy production before the hydrodynamical evolution, which may differ from the transverse shape of the proton.

In heavy nuclei, the proton and neutron shape fluctuations are usually suppressed, and the fluctuating nucleonic structure dominates many observables that are sensitive to the initial state fluctuations. However, there are processes such as exclusive vector meson production at high momentum transfer $|t|$ that probe fluctuations at small distance scales $\sim 1/|t|$ and as such, can be used to determine if the fluctuating structure of protons and neutrons is modified in the nuclear environment.

In the near future, large amounts of new data is expected to help constraining proton shape fluctuations in more detail. Detailed correlation measurements form high multiplicity proton-proton and proton-nucleus collisions at the LHC, as well as proton-light ion or light-heavy-ion collisions at RHIC will provide useful data. Photoproduction studies in ultra peripheral collisions at the LHC, and DIS experiments in a next generation Electron Ion Collider in the US~\cite{Boer:2011fh,Accardi:2012qut,Aschenauer:2017jsk}, Europe~\cite{AbelleiraFernandez:2012cc} or China~\cite{Chen:2018wyz}  will provide vast amounts of precise data probing the structure of protons and nuclei with high precision.  

On the theory side, developments need to stay in pace with the experimental advances. Existing Monte Carlo event generators, such as Sar$t$re~\cite{Toll:2012mb,Toll:2013gda} which simulates exclusive vector meson production at the EIC, could be updated to include nucleon shape fluctuations. 
More generally, there are two important paths for theory developments. First, perturbative calculations need to be developed to next-to-leading order accuracy in the QCD coupling $\alpha_s$ to achieve precision level (see  e.g. Refs.~\cite{Balitsky:2008zza,Balitsky:2010ze,Balitsky:2013fea,Kovner:2013ona,Lappi:2015fma,Lappi:2016fmu,Boussarie:2016bkq,Hanninen:2017ddy,Ducloue:2017ftk,Beuf:2017bpd,Ducloue:2019ezk,Kang:2019ysm} for recent developments in the Color Glass Condensate framework). Additionally, new more differential observables are needed to probe the proton and nuclear structure in more detail. For example, dijet production at high energies is a promising observable in ultraperipheral collisions at the LHC~\cite{ATLAS:2017kwa} and in a future Electron-Ion Collider~\cite{Dumitru:2018kuw}, as the second momentum vector provides an additional handle to the geometry and potentially enables more precise extraction of the proton and nuclear shape fluctuations. It may even be possible to extract properties of the Wigner distribution that describes how partons are distributed both as a function of transverse coordinate and transverse momentum~\cite{Hatta:2016dxp,Hagiwara:2017fye,Salazar:2019ncp,Mantysaari:2019csc,Mantysaari:2019hkq}. The effect of nucleon substructure fluctuations on multi particle interactions (MPIs) should also be studied, as these processes can potentially provide complementary constraints on the shape fluctuations.

\section*{Acknowledgements}
This work was supported by the Academy of Finland project 314764. I thank B. Schenke for useful comments and T. Lappi for a careful reading of the manuscript.

\bibliography{refs}
\bibliographystyle{tes-2modlong}

\end{document}